\documentclass[aps,pre,twocolumn,showpacs,amsmath,amssymb]{revtex4}
\pdfoutput=1

\usepackage{graphicx}
\usepackage{amsthm}

\DeclareMathOperator{\nbg}{nbg} 
\newtheorem{defn}{Definition}
\newtheorem{lemm}{Lemma}
\newtheorem{ther}{Theorem}
\newtheorem{prop}{Proposition}
\newtheorem{coro}{Corollary}

\begin{document}
\title{A stitch in time: Efficient computation of genomic DNA melting bubbles}
\author{Eivind T\o stesen} 
\email[Email: ]{eivindto@math.uio.no}
\affiliation{Department of Tumor Biology, Norwegian Radium Hospital, N-0310 Oslo, Norway,
and Department of Mathematics, University of Oslo, N-0316 Oslo, Norway}
\date{\today}

\begin{abstract}
Background: 
It is of biological interest to make genome-wide predictions of the locations
of DNA melting bubbles using statistical mechanics models.
Computationally, this poses the challenge that
a generic search through all combinations of bubble starts and ends
is quadratic.

Results:
An efficient algorithm is described, which shows that the time complexity of the
task is O(NlogN) rather than quadratic.
The algorithm exploits that bubble lengths may be limited,
but without a prior assumption of a maximal bubble length.
No approximations, such as windowing, have been introduced
to reduce the time complexity.
More than just finding the bubbles, the algorithm produces a stitch profile,
which is a probabilistic graphical model of bubbles and helical regions.
The algorithm applies a probability peak finding method 
based on a hierarchical analysis
of the energy barriers in the Poland-Scheraga model.

Conclusions:
Exact and fast computation of genomic stitch profiles is thus feasible.
Sequences of several megabases have been computed,
only limited by computer memory.
Possible applications are the genome-wide comparisons of bubbles
with promotors, TSS, viral integration sites, and other melting-related regions.
\end{abstract}

\maketitle

\section{Background}
Models of DNA melting make it possible
to compute what regions that are single-stranded (ss)
and what regions that are double-stranded (ds).
Based on statistical mechanics, such model predictions are probabilistic by nature.
Bubbles or single-stranded regions
play an essential role in fundamental biological processes,
such as transcription, replication, viral integration, repair, recombination,
and in determining chromatin structure
\cite{calladine,sumner}.
It is therefore interesting to apply DNA melting models to genomic DNA sequences,
although the available models so far are limited to \emph{in vitro} knowledge.
Genomic applications began around 1980
\cite{tong1979,wada},
and have been gaining momentum over the years
with the increasing availability of sequences, faster computers,
and model development.
It has been found that predicted ds/ss boundaries often are located at or very close to
exon-intron junctions, the correspondence being stronger in some genomes than others
\cite{GrahamJ.King09111993,Yeragenea,Yerageneb,Yerafizz},
which suggested a gene finding method \cite{Yeraproof}.
In the same vein, comparisons of actin cDNA melting maps in animals, plants, and fungi
suggested that intron insertion could have target
the sites of such melting fork junctions in ancient genes
\cite{carlon:178101,carlon:051916}.
In other studies, bubbles in promotor regions were computed
to test the hypothesis that the stability of the double helix
contributes to transcriptional regulation
\cite{choi,erp:218104,benham:059801,erp:059802,choi:239801,erp:239802}.
Bubbles induced by superhelicity have also been found to correlate
with replication origins as well as promotors
\cite{8385354,15289476,AkBenham,BenhamJMB96}.
In addition to the testing of specific hypotheses,
a strategy has been to provide whole genomes with annotations
of their melting properties
\cite{16381890,liu2007}.
Combined with all other existing annotations,
such melting data allow exploratory data mining
and possibly to form new hypotheses
\cite{gij}.
For example, the human genomic melting map was made available,
compared to a wide range of other annotations,
and was shown to provide more information than the local GC content
\cite{liu2007}.

In the genomic studies, various melting features
have proved to be of particular interest.
These include the bubbles and helical regions, bubble nucleation sites,
cooperative melting domains, melting fork junctions,
breathers, sites of high or low stability, and SIDD sites.
Most often we want to know their locations, but additional information
is sometimes useful, such as probabilities, dynamics, stabilities, and context.
DNA melting models based on statistical mechanics are powerful tools
for calculating such properties, especially those models that can 
be solved by dynamical programming in polynomial time.
For many features of interest, however, algorithms remain to
be developed to do such predictions.
The existing melting algorithms typically produce melting profiles
of some numerical quantity for each sequence position.
The prototypical example is Poland's probability profile \cite{poland},
but also profiles of melting temperatures (melting maps), free energies or
other quantities are computed per basepair.
The result can be plotted as a curve, while the wanted features
often have the format of regions, junctions and other sites.
Some genomics data mining tools also require data in these formats
rather than curves.
As a remedy, melting profiles have been subjected to
\emph{ad hoc} post-processing methods to extract the wanted features,
such as segmentation algorithms \cite{liu2007},
thresholding \cite{16381890},
and relying on the eye through visualization \cite{Yerafizz,carlon:051916}.

\begin{figure*}
 \includegraphics[width=17.6cm]{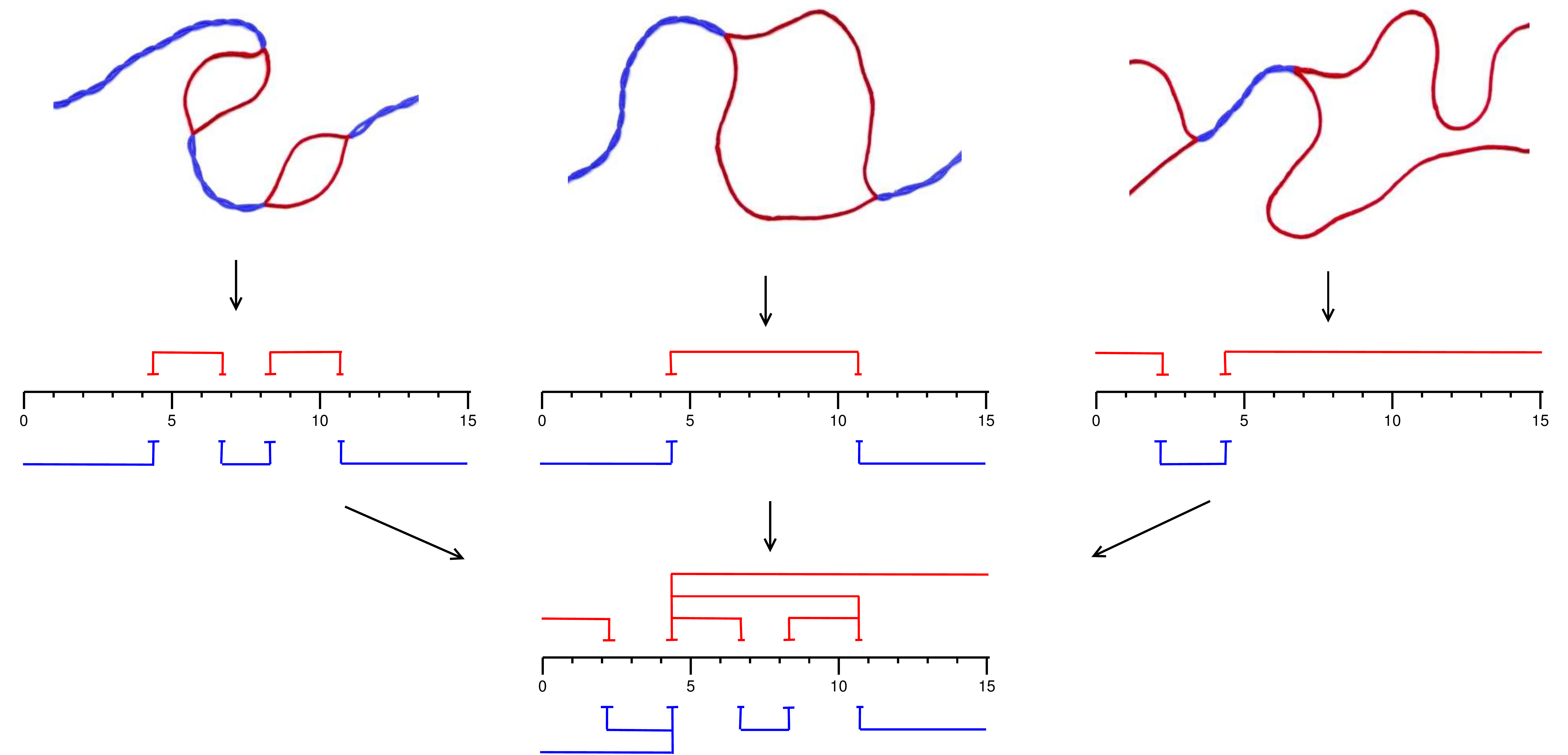}%
 \caption{
What is a stitch profile diagram?
At the top are sketched three alternative DNA conformations at the same temperature.
In the middle diagrams, the sequence location of each helical region (blue)
and each bubble or single-stranded region (red) is represented by a stitch.
At the bottom, the three ``rows of stitches'' are merged into a stitch profile diagram.
\label{whatis}
}
\end{figure*}

In previous work, we developed an algorithm
that identifies regions of four types:
helical regions, bubbles (internal loops), and unzipped 5' and 3' end regions (tails)
\cite{tostesen:061922,bip2003,narweb}.
The algorithm produces a \emph{stitch profile}, which is
a probabilistic graphical model of DNA's conformational space.
A stitch profile contains a set of regions of the four types.
Each region is called a \emph{stitch},
because of the way they can be connected in paths.
The stitch profile algorithm computes the location (start and end)
of each stitch and the probability of that region being in the
corresponding state (ds or ss) at the specified temperature.
A stitch profile can be plotted in a \emph{stitch profile diagram},
as illustrated in Fig.~\ref{whatis}.
The location of a bubble or helix stitch is not given as
a precise coordinate pair $(x,y)$, but rather as a pair of
ds/ss boundaries with fuzzy locations. For each ds/ss boundary,
the range of thermal fluctuations is computed and given as an interval.
A stitch profile indicates a number of alternative configurations,
both optimal and suboptimal, as illustrated in Fig.~\ref{whatis}.
In contrast, a melting map would indicate the single configuration at each temperature,
in which each basepair is in its most probable state.

A stitch profile thus provides some features, e.g. bubbles, that would be of
interest in genomic analyses.
However, the previously described algorithm for computing stitch profiles
\cite{tostesen:061922} has time complexity $O(N^2)$.
Genomics studies often require faster algorithms,
both to compute long sequences and to compute
many sequences. In this paper, therefore, an efficient
stitch profile algorithm with time complexity $O(N\log N)$
is described, and the prospects of
computing genomic stitch profiles are discussed.
The original algorithm \cite{tostesen:061922} is referred to as Algorithm 1,
while the new algorithm is referred to as Algorithm 2.

The reduction in time complexity has been achieved without introducing
any approximation or simplification such as windowing. The usual tradeoff between
speed and precision is therefore not involved here. The output
of Algorithm 2 is not of a lower quality, but identical to Algorithm 1's output.
Algorithm 1 was simply inefficient.
However, it was not obvious that this problem has time complexity $O(N\log N)$,
which is the same as computing melting profiles with
the Poland-Fixman-Freire algorithm \cite{FF77}.
It would appear that the stitch profile had greater complexity,
for example, that the search for all bubble starts and ends would be quadratic.
On the other hand, we know that bubbles may be small compared to
the sequence length.
Algorithm 2 detects such circumstances in an adaptive way,
without assuming a maximal bubble length.
\section{Methods}
The proper way of computing DNA conformations,
as well as other macromolecular structures,
is to consider a \emph{rugged landscape}
\cite{10.1038/nsb0197-10,10.1038/29487}.
As an abstract mathematical function, a landscape
applies to widely different complex systems,
for example, fitness landscapes in evolutionary biology
for defining populations and species.
The ruggedness implies many local maxima and minima on many levels.
In optimization, the task would be to avoid
all the ``false'' local optima and find the global optimum.
That is not what we want.
On the contrary, we would prefer to include most of them.

A local optimum corresponds to an instantaneous conformation or \emph{microstate}
that is more fit or stable than its immediate neighbors.
However, fluctuations over time cover a larger area in the landscape around the local optimum,
which is defined as a \emph{macrostate}.
A macrostate can not simply be associated with a local
optimum, because it usually covers many local optima.
On the other hand, a local optimum may be part of different macrostates.
Fluctuations are biologically important, as they represent
stability and robustness, rather than noise and uncertainty
\cite{stelling}.
Conformations are properly represented by macrostates, not microstates. 
We want to characterize the whole landscape of DNA conformations
by a set of macrostates.

More specifically, this article
considers certain probability landscapes,
in which the probability peaks are the macrostates.
The algorithmic task is to find a set of peaks.
Automatic peak detecting is applied
in various kinds of spectroscopy (NMR), spectrometry (mass-spec),
and image segmentation (e.g. in astronomy), but these algorithms
usually do not consider any hierarchical aspects.
Hierarchical peak finding is analogous to hierarchical clustering,
which is widely used in bioinformatics.
However, our approach is closely related
to the hierarchical analyses of energy landscapes and their barriers
in studies of dynamics, metastability, and timescales
\cite{PhysRevA.38.4261,PhysRevE.51.5228,stadler5,stadler1}.
The algorithm uses a subroutine for finding
hierarchical probability peaks in one dimension,
described in the next section.
\subsection{1D peaks}
This section briefly revisits the 1D peak finding method
and the use of a nonstandard pedigree terminology \cite{tostesen:061922}.
Here is a generic formulation of the problem:
Let $p(x)$ be some probabilities (possibly marginal) defined for $x=1,\ldots,N$.
What are the peaks in $p(x)$?
The computational task is divided into two steps.
The first step is to construct a discrete tree of possible peaks,
and the second step is to select peaks by searching the tree.

\begin{figure}
 \includegraphics[width=8.6cm]{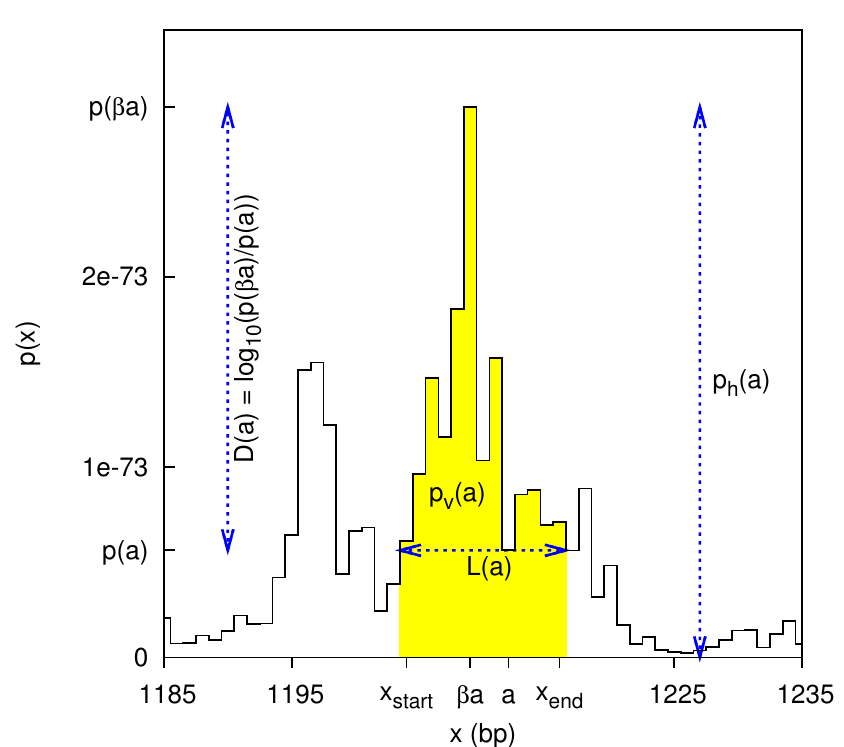}%
 \caption{
Example of a 1D peak.
This peak in $p(x)$ has
peak volume (yellow area) $p_{\text{v}}(a)=1.5\times 10^{-72}$,
while the peak height is $p_{\text{h}}(a)=2.9\times 10^{-73}$,
which is the maximum probability attained at $\beta a=1209$.
The peak location $L(a)$ is the extent
from $x_{\text{start}}=1204$ to $x_{\text{end}}=1216$,
which corresponds to the local minimum attained at $a=1212$.
The depth is $D(a)=0.711$.
\label{p1}
}
\end{figure}

To simplify the presentation, we assume that
$p(x_1)\neq p(x_2)$ if $x_1\neq x_2$.
Let $\Psi$ be the set of $x$-values, where $p(x)$ has local minima and maxima.
We associate a possible peak with each element $a\in \Psi$.
If $a$ is a local minimum, the peak is defined
as illustrated in Fig.~\ref{p1}.
The \emph{peak location} is the extent on the $x$-axis,
$L(a)=[x_{\text{start}}(a),x_{\text{end}}(a)]$,
defined as the largest interval including $a$ in which $p(x)\geq p(a)$.
The \emph{peak width} is the size of $L(a)$,
$p_{\text{w}}(a)=x_{\text{end}}(a)-x_{\text{start}}(a)+1$.
The \emph{peak volume} is the probability summed over
the location, $p_{\text{v}}(a)=\sum_{x\in L(a)}p(a)$.
The peak's \emph{bottom} $\beta a=\arg \max_{x \in L(a)} p(x)$
is the $x$-value where $p$ attains its maximum.
(The term ``bottom'' originates from the corresponding energy landscape
picture, but it is the position of the peak's top.)
The \emph{peak height} is $p_{\text{h}}(a)=p(\beta a)$.
The peak's \emph{depth} is $D(a)=\log_{10}\frac{p(\beta a)}{p(a)}$.
We also associate a possible peak with each local maximum $a\in \Psi$,
namely the spike itself:
$L(a)=[a,a]$, $p_{\text{w}}(a)=1$, $\beta a=a$,
$p_{\text{v}}(a)=p_{\text{h}}(a)=p(a)$, and $D(a)=0$.

While peaks may be high, it is a more defining characteristic that
they are wide.
A peak is produced by the fluctuations in $x$,
rather than disturbed by them.
For each local maximum, there are many possible peaks.
Therefore, a peak can not be identified with its bottom.
Instead, we use the elements in $\Psi$ as unique identifiers of peaks.
The location of a peak is $L(a)$, not the bottom position $\beta a$,
and the size of a peak is the peak volume, not the peak height.
However, for the second type of peaks (the maxima),
the peak location reduces to the bottom
and the peak volume reduces to the peak height.

The set $\Psi$ of possible peaks is hierarchically ordered.
A binary tree is defined by the set inclusion order
on the set of peak locations.
For each pair $a,a'\in\Psi$, either $L(a)\subseteq L(a')$,
or $L(a)\supseteq L(a')$, or they are disjoint.
The branching corresponds to each local minimum $a$
dividing the peak into two subpeaks, see Fig.~\ref{p1}, just as
a barrier or a watershed or a saddle point divides two valleys or lakes
in a landscape
\cite{PhysRevA.38.4261,stadler5,stadler1}.
The global minimum is the \emph{root} node $\rho$ of the tree.
The local maxima are the leaf nodes of the tree.
Each $a\in\Psi$ has at most three edges, one towards the root and
two away from the root.
Each $a\neq \rho$ has an edge towards the root
that connects to the \emph{successor} $\sigma a$.
Each successor has an increased depth: $D(\sigma a)\geq D(a)$.
And each local minimum $a$ has two edges away from the root
that connect to two \emph{ancestors}.
The highest peak of the two ancestors is the \emph{father} $\pi a$
and the other is the \emph{mother} $\mu a$, i.e., they are distinguished by
$p_{\text{h}}(\pi a)>p_{\text{h}}(\mu a)$. 
A left-right distinction between the two is not used.
The notation $\sigma^{n}a$ means the successor taken $n\geq 0$ times,
where $\sigma^{0}a=a$. 
Each $a$ has a \emph{set of successors} $\Sigma (a)$ defined as the path from
$a$ to the root: $a,\sigma a,\sigma ^{2}a,\ldots,\rho$.
Each $a$ also has a \emph{set of ancestors} $\Delta (a)$ defined by
$a'\in\Delta (a) \Leftrightarrow a\in\Sigma (a')$.
The set $\Delta (a)$ is the subtree that has $a$ as its root node.
A bottom is typically shared by several peaks.
For example, a peak has the same bottom as its father, $\beta a=\beta \pi a$,
but not the same as its mother, $\beta a\neq \beta \mu a$.
Each $a$ has a \emph{paternal line} $\Pi (a)$, defined as the set
of all nodes that share $a$'s bottom.
$\Pi (a)$ is also the path including $a$ connected by fathers
that ends at $\beta a$.
The beginning of the path, called the \emph{full} node $\varphi a$,
is either a mother or the root.
The paternal lines establish a one-to-one correspondence between
the set of maxima (i.e. bottoms) and the set of mothers including the root.

Having established a hierarchy $\Psi$ of possible peaks,
the second step is to select among them. The selection
applies two independent criteria, each controlled by an
input parameter:
the \emph{maximum depth} $D_{\text{max}}$
and the \emph{probability cutoff} $p_{\text{c}}$.
The first criterion is that $a$ is a \emph{1D peak}
according to the following definition.
\begin{defn}
Let $D_{\text{max}}$ be the maximum depth of peaks.
Then $a\in \Psi$ is a \emph{1D peak} if
\begin{enumerate}
\item $D(a)<D_{\text{max}}$,
\item $D(\sigma a)\geq D_{\text{max}}$ or $a= \rho$.
\end{enumerate}
\label{maxdeep1D}
\end{defn}
The second criterion is that $p_{\text{v}}(a)\geq p_{\text{c}}$.
The first criterion is invoked by
using the \textsc{maxdeep} subroutine \cite{tostesen:061922},
which returns the set $P$ of all 1D peaks.
The second criterion is subsequently invoked by calculating the peak volume 
of each $a\in P$ and comparing with the probability cutoff.
\subsection{Bubbles and helical regions}
The stitch profile algorithm is separate from the
statistical mechanical DNA melting model.
The only interface to the underlying model
is by calling the following probability functions:
\begin{eqnarray}
p_{\text{right}}(x)
&=& P(
\text{\ldots XX}
\stackrel{x}{1}
\underbrace{0\cdots 0}_{\text{unzipped}}
\text{--3'}
) \label{pright},\\
p_{\text{left}}(y)
&=& P(
\text{5'--}
\underbrace{0\cdots 0}_{\text{unzipped}}
\stackrel{y}{1}
\text{XX\ldots}
) \label{pleft},\\
p_{\text{bubble}}(x,y)
&=& P(
\text{\ldots XX}
\stackrel{x}{1}
\underbrace{0\cdots 0}_{\text{bubble}}
\stackrel{y}{1}
\text{XX\ldots}
) \label{pbubble},\\
p_{\text{helix}}(x,y)
&=& P(
\text{\ldots XX0}
\underbrace{\stackrel{x}{1}\cdots \stackrel{y}{1}}_{\text{helix}}
\text{0XX\ldots}
) \label{phelix},\\
p_{\text{helix}}(x,N)
&=& P(
\text{\ldots XX0}
\underbrace{\stackrel{x}{1}\cdots 1}_{\text{zipped}}
\text{--3'}
) \label{phelix_x},\\
p_{\text{helix}}(1,y)
&=& P(
\text{5'--}
\underbrace{1\cdots \stackrel{y}{1}}_{\text{zipped}}
\text{0XX\ldots}
) \label{phelix_y}.
\end{eqnarray}
In these equations,
1 is a bound basepair (helix),
0 is a melted basepair (coil),
X is either 0 or 1,
and the sequence positions $x$ and/or $y$ are indicated.

In addition to these, the stitch profile algorithm calls methods
for adding these probabilites (peak volumes) and for computing
upper bounds on such probability sums.
This means that it is easy to change or replace the underlying model.
In this article, the Poland-Scheraga model with Fixman-Freire loop entropies
is used \cite{bip2003}, but in principle, other DNA melting models could be used,
or even models that include secondary structure
\cite{zukerhybrid}.

This article discusses how to efficiently compute bubble stitches and
helix stitches only. The 5' and 3' tail stitches
are efficiently computed as in Algorithm 1
\cite{tostesen:061922}.
Each bubble stitch corresponds to a peak in
the bubble probability function in Eq.~(\ref{pbubble}).
And each helix stitch corresponds to a peak in
the helix probability function in Eq.~(\ref{phelix}).
These two probability functions and their peaks are two dimensional,
so the 1D peak finding method does not directly apply.
However, the 1D peak analysis can be performed for each
of the other four probability functions
[Eqs.~(\ref{pright}), (\ref{pleft}), (\ref{phelix_x}), and (\ref{phelix_y})].
Using Eq.~(\ref{pright}), a binary tree $\Psi_{\text{x}}$ and a set
of 1D peaks $P_{\text{x}}$ is computed,
and using Eq.~(\ref{pleft}), a binary tree $\Psi_{\text{y}}$ and a set
of 1D peaks $P_{\text{y}}$ is computed.
The probability cutoff is not invoked here.
These two tree structures with their 1D peaks are then further processed,
as described in the following two sections, to obtain
the bubble stitches.
Likewise,
using Eq.~(\ref{phelix_x}), a binary tree $\Psi_{\text{x}}$ and a set
of 1D peaks $P_{\text{x}}$ is computed,
and using Eq.~(\ref{phelix_y}), a binary tree $\Psi_{\text{y}}$ and a set
of 1D peaks $P_{\text{y}}$ is computed.
These are used similarly to obtain the helix stitches.
This division of labor also indicates an obvious parallelization
of the algorithm using two or four processors.
Parallelism was not implemented in this study, however.
\subsection{2D peaks}
The goal of this section is to define 2D peaks and
to prove the key result that some 2D peaks
are simply the Cartesian product of two 1D peaks.
But not all 2D peaks have this property, making it a nontrivial result.
This is expressed in Theorem \ref{s12&13}.

Theorem \ref{s12&13} also indicates a convenient way of computing all 2D peaks,
on which Algorithm 2 is directly based. Theorem \ref{s12&13}
shows that Algorithm 2's computation of stitch profiles is exact, that is,
complying strictly with the mathematical definition of 2D peaks.
The proof is therefore important for the validation of Algorithm 2.
While Theorem \ref{s12&13} is the primary goal,
we also prove Theorem \ref{s17&18} which similarly
provides validation of Algorithm 1.
But more importantly, a comparison of the two theorems
gives more insight in both algorithms.

A \emph{frame} is a pair $(a,b)\in \Psi_{\text{x}}\times \Psi_{\text{y}}$.
A frame also refers to the corresponding box $L(a)\times L(b)$ in the $xy$-plane.
A frame $(a,b)$ is \emph{contained inside} another frame $(a',b')$,
if $L(a)\times L(b)\subset L(a')\times L(b')$,
that is, if $a'\in \Sigma (a)$ and $b'\in \Sigma (b)$.
The \emph{root frame} is
$(\rho _{\text{x}},\rho _{\text{y}})$.
A frame $(a,b)$ is \emph{nonroot} if $(a,b)\neq (\rho _{\text{x}},\rho _{\text{y}})$.
A frame $(a,b)$ is a \emph{bottom frame} if
$(a,b)=(\beta a,\beta b)$ and
it is \emph{nonbottom} if $(a,b)\neq (\beta a,\beta b)$.
The \emph{depth} of a frame $(a,b)$ is $D(a,b)= \max \{D(a),D(b)\}$.
From this definition, we immediately get
\begin{equation}
D(a,b)<D_{\text{max}} \Leftrightarrow D(a)<D_{\text{max}} \text{ and } D(b)<D_{\text{max}} .
\label{s0}
\end{equation}
To simplify the presentation, we assume that for all frames: $D(a)\neq D(b)$.
\begin{defn}
The \emph{successor} of a nonroot frame $(a,b)$ is
\begin{equation}
\sigma(a,b)=\left\{
 \begin{array}{l}
  (\sigma a,b) \text{ if } D(\sigma b)>D(\sigma a) \text{ or } b=\rho_{\text{y}}\\
  (a,\sigma b) \text{ if } D(\sigma a)>D(\sigma b) \text{ or } a=\rho_{\text{x}}
 \end{array}
\right.
\end{equation}
A successor of the root frame does not exist.
\label{sigma}
\end{defn}
Having defined the depth and the successor, what is the depth of a successor?
\begin{prop}
For every nonroot $(a,b)$, $D(\sigma (a,b))\geq D(a,b)$.
\label{s7}
\end{prop}
\begin{proof}
For $\sigma (a,b)=(\sigma a,b)$, $\max \{D(\sigma a),D(b)\}\geq \max \{D(a),D(b)\}$
because $D(\sigma a)\geq D(a)$. Likewise for $\sigma (a,b)=(a,\sigma b)$.
\end{proof}

\begin{defn}
A frame $(a,b)$ is \emph{$\sigma$-above} if
\begin{enumerate}
\item $D(\sigma a)>D(b)$ or $a= \rho _{\text{x}}$,
\item $D(\sigma b)>D(a)$ or $b= \rho _{\text{y}}$.
\end{enumerate}
\label{sigma-above}
\end{defn}
The term ``$\sigma$-above'' is a mnemonic for the two inequalities
in the definition.
The set of all frames that are $\sigma$-above is called \emph{the frame tree}.
While Prop.~\ref{s7} only sets a lower bound on the depth of a successor,
we can write the actual value for $\sigma$-above frames:
\begin{prop}
If $(a,b)$ is nonroot and $\sigma$-above, then
\begin{equation}
D(\sigma(a,b))=\left\{
 \begin{array}{l}
  D(\sigma a) \text{ if } \sigma(a,b)=(\sigma a,b)\\
  D(\sigma b) \text{ if } \sigma(a,b)=(a,\sigma b).
 \end{array}
\right.
\end{equation}
Furthermore, $D(\sigma(a,b))=min \{D(\sigma a),D(\sigma b)\}$
if both $a\neq \rho_{\text{x}}$ and $b\neq \rho_{\text{y}}$.
\label{s1&3}
\end{prop}
\begin{proof}
If $\sigma(a,b)=(\sigma a,b)$, then $a\neq \rho _{\text{x}}$ and
$\max\{D(\sigma a),D(b)\}=D(\sigma a)$
by Def.~\ref{sigma-above}.
If, furthermore, $b\neq \rho_{\text{y}}$, then 
$D(\sigma (a,b))=D(\sigma a)<D(\sigma b)$ by Def.~\ref{sigma}.
Likewise if $\sigma(a,b)=(a,\sigma b)$.
\end{proof}
By repeatedly taking the successor, we eventually end up at the root frame in, say, $R$ steps.
$\Sigma (a,b)$ is the \emph{sequence of successors} of $(a,b)$, i.e., the sequence
$\{\sigma^{n}(a,b)\}_{0}^{R}$ that begins at $(a,b)$ and ends at the root frame.
Alternatively, $\Sigma (a,b)$ is defined as the \emph{set of successors}, i.e.,
the set of such sequence elements. What if we want to exclude $(a,b)$ from
$\Sigma (a,b)$? That can be written as $\Sigma (\sigma (a,b))$.

If $(a,b)$ is not $\sigma$-above, then its sequence of successors
takes the shortest path to a $\sigma$-above frame, or put another way:
\begin{prop}
If $a'\in \Sigma (a)$, $b'\in \Sigma (b)$ and $(a',b')$ is $\sigma$-above,
then $(a',b')\in \Sigma (a,b)$.
\label{s20}
\end{prop}
\begin{proof}
All elements in both $\Sigma (a)$ and $\Sigma (b)$
are visited by the sequence $\Sigma (a,b)$ on its climb to the root frame.
Assume $(a',b')\notin \Sigma (a,b)$.
Then either $a'$ is passed before $b'$ is reached, or viceversa,
and we can assume that $a'$ comes first.
In other words, $a' \neq \rho _{\text{x}}$ and there is a $b'' \neq b'$
such that $b'\in \Sigma (b'')$ and $\sigma (a',b'')=(\sigma a',b'')$. 
Then $D(b')\geq D(\sigma b'')$.
By Def.~\ref{sigma}, we see that $D(\sigma b'')>D(\sigma a')$.
$(a',b')$ is $\sigma$-above, so
by Def.~\ref{sigma-above}, we see that $D(\sigma a')>D(b')$.
We arrive at the contradiction $D(b')>D(b')$.
\end{proof}
Each frame is the successor of at most four frames.
If $(a,b)=\sigma (a',b')$ then $(a',b')$ is either
$(\pi a,b)$, $(a,\pi b)$, $(\mu a,b)$, or $(a,\mu b)$.
Two of these are defined as \emph{ancestors}:
\begin{defn}
The \emph{father} of a nonbottom frame $(a,b)$ is
\begin{equation}
\pi(a,b)=\left\{
                     \begin{array}{l}
                     (\pi a,b) \text{ if } D(a)>D(b)\\
                     (a,\pi b) \text{ if } D(a)<D(b)
 		     \end{array}
		\right.
\label{pi}
\end{equation}
The \emph{mother} of a nonbottom frame $(a,b)$ is
\begin{equation}
\mu(a,b)=\left\{
                     \begin{array}{l}
                     (\mu a,b) \text{ if } D(a)>D(b)\\
                     (a,\mu b) \text{ if } D(a)<D(b).
 		     \end{array}
		\right.
\label{mu}
\end{equation}
Fathers and mothers of bottom frames do not exist.
\label{pi&mu}
\end{defn}
Each father or mother can have its own father and mother, and so on.
The \emph{set of ancestors} $\Delta (a,b)$ is the binary subtree
defined recursively by:
(1) $(a,b)\in \Delta (a,b)$.
(2) If nonbottom $(a',b')\in \Delta (a,b)$
then $\pi (a',b')\in \Delta (a,b)$
and $\mu (a',b')\in \Delta (a,b)$.

The next proposition shows that being $\sigma$-above is propagated
by $\sigma$, $\pi$, and $\mu$:
\begin{prop}
Let $(a,b)$ be $\sigma$-above.
\begin{enumerate}
\item If $(a',b')\in \Sigma (a,b)$ then $(a',b')$ is $\sigma$-above.
\item If $(a',b')\in \Delta (a,b)$ then $(a',b')$ is $\sigma$-above.
\end{enumerate}
\label{s16}
\end{prop}
\begin{proof}
(i): First, we show that $\sigma(a,b)$ is $\sigma$-above:
If $\sigma(a,b)=(\sigma a,b)$, then
Def.~\ref{sigma} implies the second condition:
$D(\sigma b)>D(\sigma a)$ or $b=\rho_{\text{y}}$.
And $(a,b)$ is $\sigma$-above which by Def.~\ref{sigma-above}
implies the first condition:
$D(\sigma^{2} a)>D(\sigma a)>D(b)$ or $\sigma a=\rho_{\text{x}}$.
Similarly, $\sigma(a,b)=(a,\sigma b)$ is shown to be $\sigma$-above.
The proof is completed by induction.

(ii): First, we show that $\pi(a,b)$ is $\sigma$-above:
If $\pi(a,b)=(\pi a,b)$, then
Eq.~(\ref{pi}) implies the first condition:
$D(\sigma \pi a)=D(a)>D(b)$ or $\pi a=\rho_{\text{x}}$.
And $(a,b)$ is $\sigma$-above which by Def.~\ref{sigma-above}
implies the second condition:
$D(\sigma b)>D(a)>D(\pi a)$ or $b=\rho_{\text{y}}$.
Similarly, $\pi(a,b)=(a,\pi b)$ and $\mu(a,b)$ are shown to be $\sigma$-above.
The proof is completed by induction.
\end{proof}
Successors are the inverse of fathers and/or mothers
for $\sigma$-above frames only:
\begin{prop}
If $(a,b)$ is nonbottom and nonroot,
the following statements are equivalent:
\begin{enumerate}
\item $(a,b)$ is $\sigma$-above
\item $\sigma \pi(a,b)=(a,b)$
\item $\sigma \mu(a,b)=(a,b)$
\item $\pi \sigma(a,b)=(a,b)$ or $\mu \sigma(a,b)=(a,b)$
\end{enumerate}
\label{s21}
\end{prop}
\begin{proof}
$(i)\Leftrightarrow (ii)$:
If $\pi(a,b)=(\pi a,b)$, then Eq.~(\ref{pi}) implies the first condition
that $(a,b)$ is $\sigma$-above:
$D(\sigma a)>D(a)>D(b)$ or $a=\rho_{\text{x}}$.
Then
$(a,b)$ is $\sigma$-above
$\stackrel{\text{Def.}~\ref{sigma-above}}{\Longleftrightarrow}$
$D(\sigma b)>D(a)=D(\sigma \pi a)$ or $b= \rho _{\text{y}}$
$\stackrel{\text{Def.}~\ref{sigma}}{\Longleftrightarrow}$
$\sigma (\pi a,b)=(\sigma \pi a,b)\Longleftrightarrow \sigma \pi(a,b)=(a,b)$.
If $\pi(a,b)=(a,\pi b)$, the equivalence is shown similarly.

$(i)\Leftrightarrow (iii)$:
Replace $\pi$ by $\mu$ in the above.

$(i)\Leftrightarrow (iv)$:
If $\sigma(a,b)=(\sigma a,b)$, then Def.~\ref{sigma} implies the second condition
that $(a,b)$ is $\sigma$-above:
$D(\sigma b)>D(\sigma a)>D(a)$ or $b= \rho _{\text{y}}$.
Then
$(a,b)$ is $\sigma$-above
$\stackrel{\text{Def.}~\ref{sigma-above}}{\Longleftrightarrow}$
$D(\sigma a)>D(b)$
$\stackrel{\text{Def.}~\ref{pi&mu}}{\Longleftrightarrow}$
$\pi (\sigma a,b)=(\pi \sigma a,b)$ or $\mu (\sigma a,b)=(\mu \sigma a,b)$
$\Longleftrightarrow$
$\pi \sigma(a,b)=(a,b)$ or $\mu \sigma(a,b)=(a,b)$.
If $\sigma(a,b)=(a,\sigma b)$, the equivalence is shown similarly.
\end{proof}
Accordingly, there is an ``inverse''  relationship between the sets of successors and ancestors:
\begin{prop}
$(a',b')$ is $\sigma$-above and $(a,b)\in \Sigma (a',b')$
\emph{iff}
$(a,b)$ is $\sigma$-above and $(a',b')\in \Delta (a,b)$.
\label{s22}
\end{prop}
\begin{proof}
$(a,b)\in \Sigma (a',b')$
implies a path of successors from $(a',b')$ to $(a,b)$. 
Prop.~\ref{s16} shows that all elements in the path are $\sigma$-above.
Prop.~\ref{s21}(iv) applied to each step in the path
gives an opposite path of ancestors.

Conversely, $(a',b')\in \Delta (a,b)$
implies a path of ancestors from $(a,b)$ to $(a',b')$. 
Prop.~\ref{s16} shows that all elements in the path are $\sigma$-above.
Prop.~\ref{s21}(ii) and (iii) applied to each step in the path
gives an opposite path of successors.
\end{proof}
It follows from Prop.~\ref{s22} that the frame tree
is equal to the binary tree $\Delta (\rho _{\text{x}},\rho _{\text{y}})$, because
$(\rho _{\text{x}},\rho _{\text{y}})\in \Sigma (a',b')$ for any
$(a',b')$.
It has the same pedigree properties as $\Psi$, such as
paternal lines and $\beta \pi (a,b)=\beta (a,b)$.

\begin{figure*}
 \includegraphics[width=13cm]{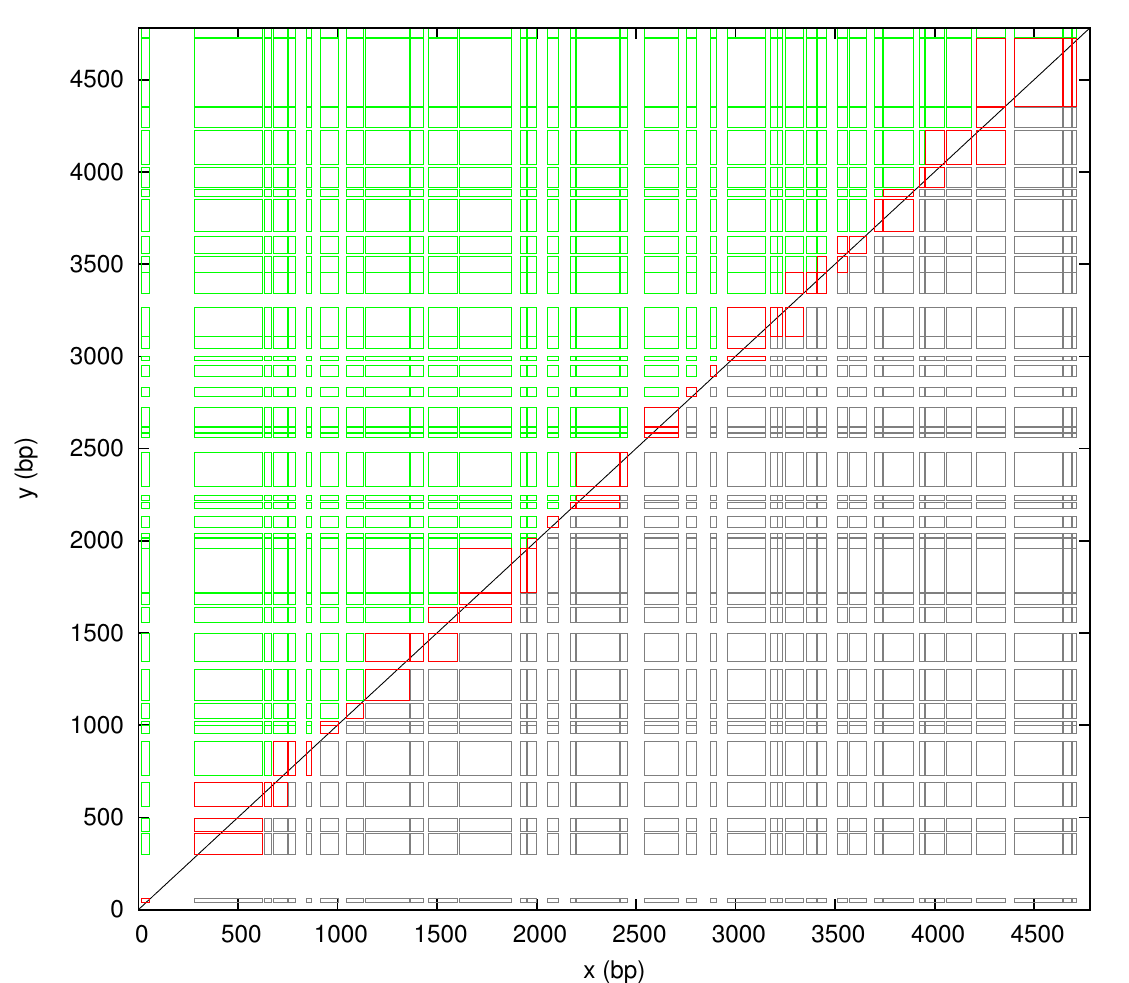}%
 \caption{
 The set $G=P_{\text{x}}\times P_{\text{y}}$ of all grid frames
plotted in the $xy$-plane. 
The grid frames are colored to distinguish those that are
above the diagonal (green),
crossing the diagonal (red), and
below the diagonal (grey),
thus illustrating the subsets $G_{\text{a}}$, $G_{\text{c}}$ and $G_{\text{b}}$, respectively.
Frames with side lengths below 20 bp are not shown to unclutter the figure. 
\label{grid}
}
 \end{figure*}

So far, we have covered ground that was already implicit in \cite{tostesen:061922},
but augmented here with proofs.
The next concept is new, however, namely the
Cartesian products of 1D peaks.
\begin{defn}
$(a,b)$ is a \emph{grid frame} if $a$ and $b$ are 1D peaks.
\end{defn}
The set of all grid frames is $G=P_{\text{x}}\times P_{\text{y}}$.
As Fig.~\ref{grid} shows,
$G$ has a grid-like ordering in the $xy$-plane.
All 1D peaks $a\in P_{\text{x}}$ have disjoint peak locations
$L(a)=[x_{\text{start}}(a),x_{\text{end}}(a)]$.
They can be indexed by $i=1,2,3,\ldots$
according to their ordering from 5' to 3' on the sequence, such that
$x_{\text{end}}(a_{i})<x_{\text{start}}(a_{i+1})$.
Likewise, the 1D peaks $b\in P_{\text{y}}$ can be indexed by $j$.
Then the grid frames form a matrix $G$
with elements $[G]_{ij}=(a_{i},b_{j})$. We use the symbol
$G$ for both the set and the matrix.

\begin{prop}
Every grid frame $(a,b)$ is $\sigma$-above. 
\label{s5}
\end{prop}
\begin{proof}
If $a\neq \rho _{\text{x}}$, then 
$D(\sigma a)\geq D_{\text{max}}$ because $a$ is a 1D peak
and $D_{\text{max}}>D(b)$ because $b$ is a 1D peak (see Def.~\ref{maxdeep1D}),
thus showing Def.~\ref{sigma-above}(i). 
Similarly, we show Def.~\ref{sigma-above}(ii).
\end{proof}

The following two lemmas show that grid frames inherit some properties from 1D peaks.
\begin{lemm}
$(a,b)$ is a grid frame \emph{iff}
\begin{enumerate}
\item $(a,b)$ is $\sigma$-above,
\item $D(a,b)<D_{\text{max}}$,
\item $D(\sigma (a,b))\geq D_{\text{max}}$ or $(a,b)$ is the root frame.
\end{enumerate}
\label{s6}
\end{lemm}
\begin{proof}
If $(a,b)$ is a grid frame, then it is $\sigma$-above by Prop.~\ref{s5} and 
Eq.~(\ref{s0}) implies $D(a,b)<D_{\text{max}}$. For nonroot $(a,b)$,
$D(\sigma (a,b))$ equals either
$D(\sigma a)$ or $D(\sigma b)$ (Prop.~\ref{s1&3}),
which is $\geq D_{\text{max}}$ because $a$ and $b$ are 1D peaks.

Conversely, Eq.~(\ref{s0}) implies $D(a)<D_{\text{max}}$. For $a=\rho _{\text{x}}$,
$a$ is then a 1D peak. For $a\neq \rho _{\text{x}}$, Prop.~\ref{s1&3}
gives
$D(\sigma a)\geq D(\sigma (a,b))\geq D_{\text{max}}$, so $a$ is a 1D peak. Similarly,
$b$ is shown to be a 1D peak.
\end{proof}

\begin{lemm}
Let $D_{\text{max}}$ be the maximum depth of peaks.
\begin{enumerate}
\item For each $a$ with $D(a)<D_{\text{max}}$, there is exactly one
1D peak $a' \in \Sigma (a)$. \label{s9}
\item For each $(a,b)$ with $D(a,b)<D_{\text{max}}$, there is exactly one
grid frame $(a',b') \in \Sigma (a,b)$. \label{s14} 
\end{enumerate}
\label{s9&14}
\end{lemm}
\begin{proof}
(\ref{s9}): The depth increases monotonically in the sequence 
$\Sigma (a)$ of successors ( $\forall n:D(\sigma^{n}a)\leq D(\sigma^{n+1}a)$).
For $D(\rho _{\text{x}})\geq D_{\text{max}}$, there is therefore a unique element $a'\neq \rho _{\text{x}}$ with
$D(a')<D_{\text{max}}$ and $D(\sigma a')\geq D_{\text{max}}$.
For $D(\rho _{\text{x}})<D_{\text{max}}$, $a'=\rho _{\text{x}}$ is a 1D peak and no other
element in $\Sigma (a)$ can fulfill Def.~\ref{maxdeep1D}(ii).

(\ref{s14}): Eq.~(\ref{s0}) gives $D(a)<D_{\text{max}}$ and $D(b)<D_{\text{max}}$.
By applying (\ref{s9}) to $a$ and $b$, we obtain a unique grid frame $(a',b')$
where $a' \in \Sigma (a)$ and $b' \in \Sigma (a)$.
$(a',b')$ is $\sigma$-above by Prop.~\ref{s5},
so $(a',b') \in \Sigma (a,b)$ by Prop.~\ref{s20}.
\end{proof}
How do we define 2D peaks?
A straightforward way would be to generalize 1D peaks
by simply rewriting Def.~\ref{maxdeep1D} in the frame tree context.
The result would be the grid frames, as we see by Lemma \ref{s6}.
However, there is more to the picture than the frame tree,
due to a further constraint to be discussed next,
which requires a more elaborate definition of 2D peaks.

In genomic annotations, a region is specified
by coordinates $x..y$, where by convention $x<y$, i.e.,
$x$ is the 5' end and $y$ is the 3' end.
We adopt the same constraint for our notation $(x,y)$ of
the instantaneous location of a bubble or helix.
In the $xy$-plane, helices are only defined for $(x,y)$
above the diagonal line $y=x$.
Bubbles have at least one melted basepair in between $x$ and $y$,
so they are only defined for $(x,y)$
above the diagonal line $y=x+1$.
Accordingly, we require that frames are above the diagonal line,
as defined in the following.
\begin{defn}
A frame $(a,b)$ is \emph{above the diagonal} if
\begin{subequations}
\begin{equation}
x_{\text{end}}(a)+1<y_{\text{start}}(b) \text{ for bubbles},
\end{equation}
\begin{equation}
x_{\text{end}}(a)<y_{\text{start}}(b) \text{ for helices}.
\end{equation}
\end{subequations}
A frame $(a,b)$ is \emph{below the diagonal} if
\begin{subequations}
\begin{equation}
x_{\text{start}}(a)+1\geq y_{\text{end}}(b) \text{ for bubbles},
\end{equation}
\begin{equation}
x_{\text{start}}(a)\geq y_{\text{end}}(b) \text{ for helices}.
\end{equation}
\end{subequations}
A frame $(a,b)$ is \emph{crossing the diagonal} if
it is neither above the diagonal nor below the diagonal.
\label{diagonal}
\end{defn}
Note: A frame that is crossing the diagonal contains
at least one point $(x,y)$ above the diagonal line, while a frame that is below the diagonal
contains no points above the diagonal line, but its upper left corner may be
on the diagonal line.
Figure \ref{grid} illustrates frames that are above, crossing and below the diagonal.

\begin{figure*}
 \includegraphics[width=13cm]{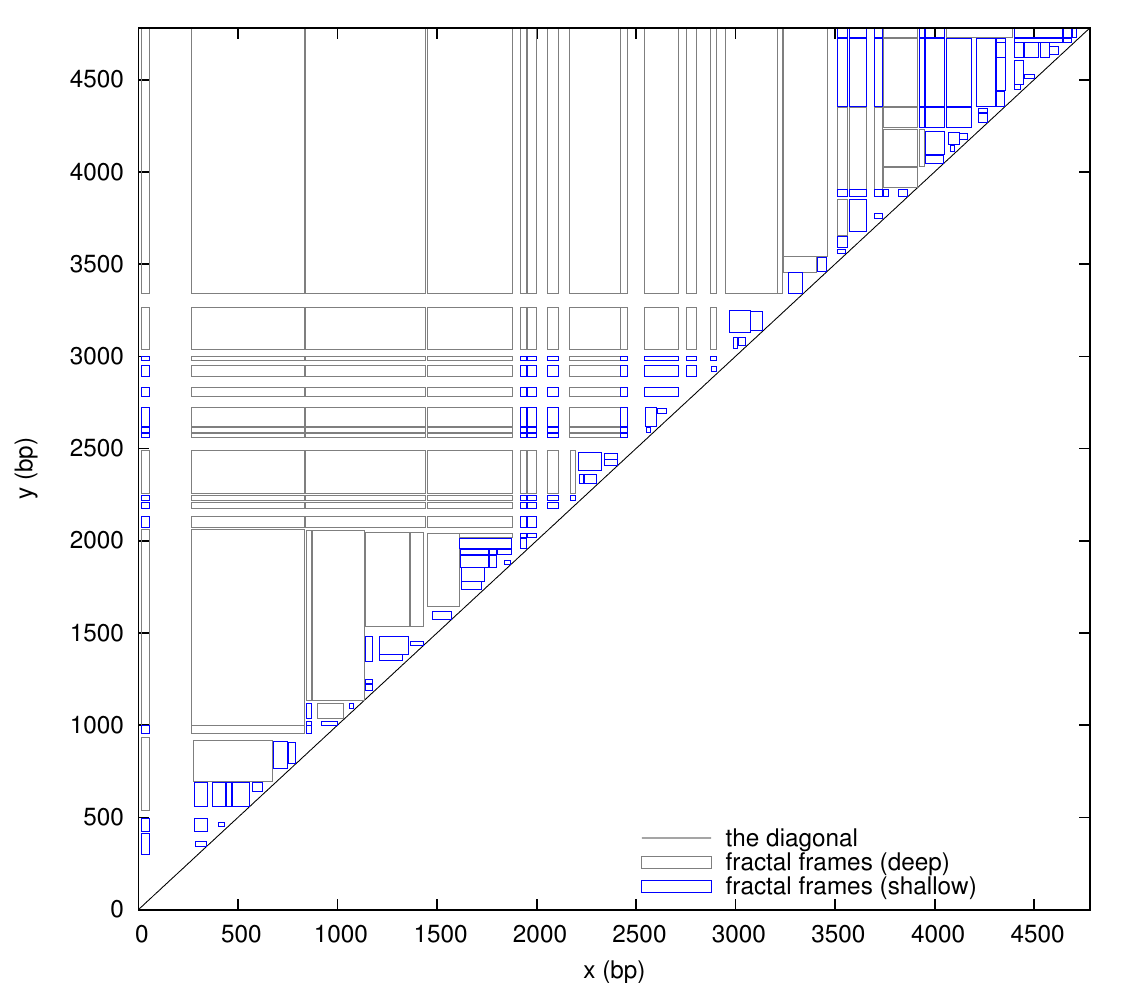}%
 \caption{
The set $F$ of all fractal frames
plotted in the $xy$-plane. 
The fractal frames $(a,b)\in F$ are colored to distinguish those with depths
$D(a,b)\geq D_{\text{max}}$ (grey) and
$D(a,b)<D_{\text{max}}$ (blue),
thus illustrating the subsets $F_{\text{d}}$ and $F_{\text{s}}$, respectively.
Frames with side lengths below 20 bp are not shown to unclutter the figure. 
\label{fractal}
}
\end{figure*}

The requirement that a frame is above the diagonal puts a constraint on its
size. This is embodied in the next concept.
\begin{defn}
The root frame is a \emph{fractal frame} if it is above the diagonal.
A nonroot frame $(a,b)$ is a \emph{fractal frame} if
\begin{enumerate}
\item $(a,b)$ is above the diagonal,
\item $\sigma (a,b)$ is crossing the diagonal,
\item $(a,b)$ is $\sigma$-above. 
\end{enumerate}
\label{treetop}
\end{defn}
The set of all fractal frames is denoted $F$.
As Fig.~\ref{fractal} shows,
fractal frames tend to be smaller the closer they are to the diagonal,
thus resembling a fractal.
For a typical fractal frame, the fluctuations in $x$ and $y$ 
are comparable in size to the length $y-x$ of the bubble or helix itself.
Indeed, the two peak locations $L(a)$ and $L(b)$ are as wide as possible,
while not overlapping each other (because the successor is crossing the diagonal).
In contrast, the fluctuations for grid frames are relatively small on average
and independent of the bubble or helix length.
\begin{lemm}
For each $\sigma$-above and above the diagonal $(a,b)$, there is exactly one
fractal frame $(a',b') \in \Sigma (a,b)$.
\label{s19}
\end{lemm}
\begin{proof}
Let $(a',b')=\sigma^{n}(a,b)$, where $n$ is the largest number for which
$\sigma^{n}(a,b)$ is above the diagonal.
$(a',b')$ is $\sigma$-above by Prop.~\ref{s16}.
For all $m>n$, frames $\sigma^{m}(a,b)$ (if they exist)
are not above the diagonal, nor below the diagonal because they contain $(a,b)$,
hence they are crossing the diagonal.
Therefore $(a',b')$ is a fractal frame.
For all $m<n$, frames $\sigma^{m}(a,b)$ (if they exist)
are above the diagonal, because they are contained in $(a',b')$. 
Therefore $(a',b')$ is the only fractal frame in $\Sigma (a,b)$.
\end{proof}
Lemma \ref{s19} is similar to Lemma \ref{s9&14}.
By Prop.~\ref{s22}, we can express both lemmas in terms of
ancestors $\Delta$ instead of successors $\Sigma$.
The lemmas then say that certain kinds of frames are organized as forests.
A \emph{forest} is a set of disjoint trees. The sets $F$ and $G$
generate two forests:
$\bigcup_{(a,b)\in G}\Delta (a,b)$ consists of the
subtrees having grid frames as root nodes.
$\bigcup_{(a,b)\in F}\Delta (a,b)$ consists of the
subtrees having fractal frames as root nodes.
By these forests, we generate from $G$ the set of all $\sigma$-above frames with $D(a,b)<D_{\text{max}}$,
and we generate from $F$ the set of all $\sigma$-above frames above the diagonal.

\begin{figure*}
 \includegraphics[width=13cm]{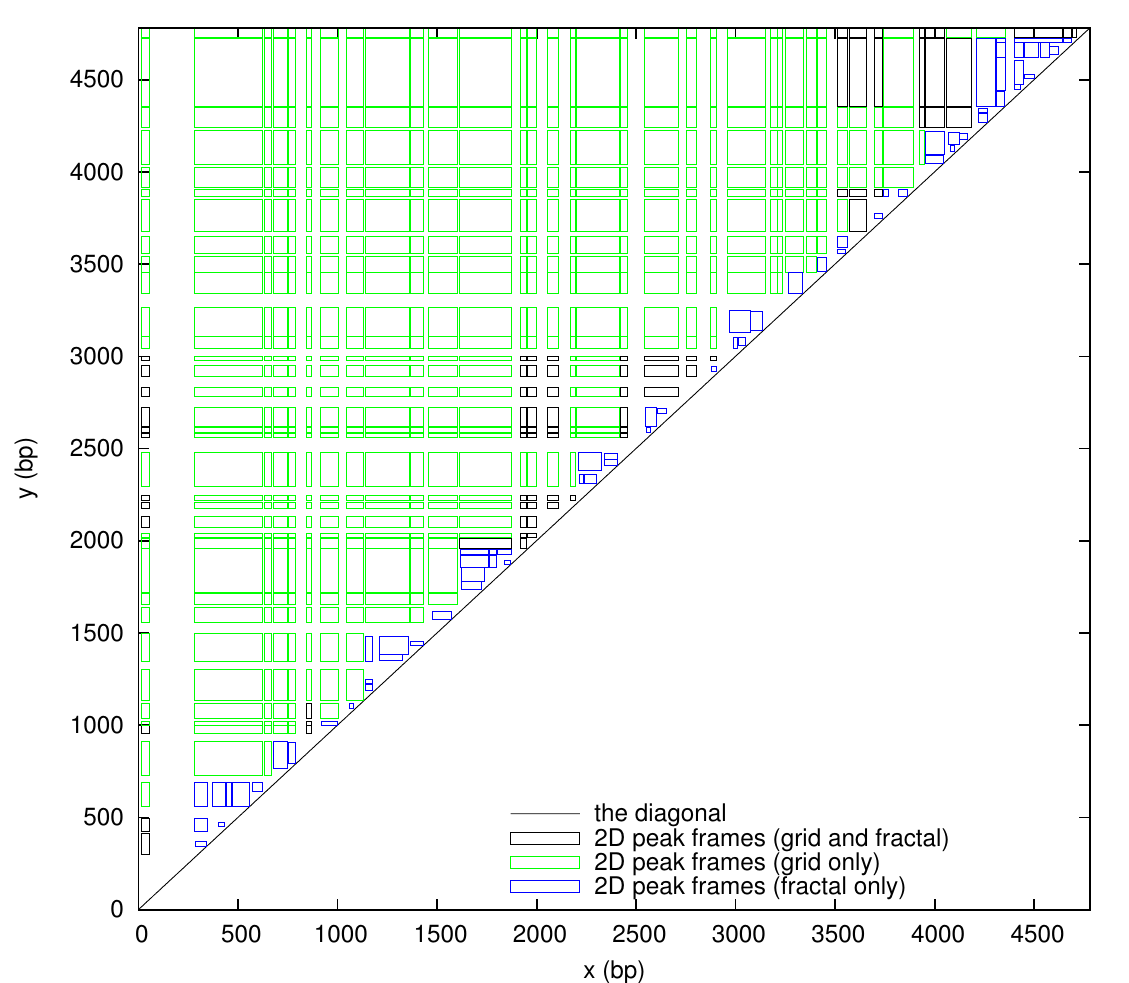}%
 \caption{
The set $P$ of all 2D peaks plotted in the $xy$-plane. 
The 2D peak frames are colored to distinguish those that are
fractal frames (blue),
fractal frames and grid frames (black), or
grid frames (green),
thus illustrating the subsets $P_{\text{F}}$, $P_{\text{FG}}$ and $P_{\text{G}}$, respectively.
Frames with side lengths below 20 bp are not shown to unclutter the figure. 
 \label{2Dpeaks}
}
 \end{figure*}

All the necessary concepts are now in place for the definition of 2D peaks.
We will not repeat the ``derivation'' of 2D peaks given in \cite{tostesen:061922},
but just recall that 2D peaks are defined with a purpose: They must capture
the extent of the actual peaks in the probability functions
$p_{\text{bubble}}(x,y)$ and $p_{\text{helix}}(x,y)$.
And they must have an interpretation in terms of fluctuations
on a given timescale.
The following definition is equivalent to the
formulation in \cite{tostesen:061922}.
\begin{defn}
Let $D_{\text{max}}$ be the maximum depth of peaks. 
A frame $(a,b)$ is a \emph{2D peak} if
\begin{enumerate}
\item $(a,b)$ is above the diagonal,
\item $(a,b)$ is $\sigma$-above,
\item $D(a,b)<D_{\text{max}}$,
\item $D(\sigma (a,b))\geq D_{\text{max}}$ or $(a,b)$ is a fractal frame.
\end{enumerate}
\label{maxdeep2D}
\end{defn}
Note: the \emph{or} in the definition is not an \emph{exclusive or}.
A 2D peak $(a,b)$ can both be a fractal frame
\emph{and} have $D(\sigma (a,b))\geq D_{\text{max}}$.
The set of all 2D peaks is denoted $P$ and is illustrated in Fig.~\ref{2Dpeaks}.

Comparing Def.~\ref{maxdeep2D} and Lemma \ref{s6}, we see that the difference
between 2D peaks and grid frames is due to the diagonal constraint:
First, the requirement that 2D peaks are above the diagonal,
and second, the possible exemption from the second inequality,
which for grid frames is being the root frame,
while for 2D peaks it is being a fractal frame.
Unlike grid frames, 2D peaks can capture events close to the diagonal
by adapting their size.

Computing the 2D peaks is at the core of the stitch profile methodology.
The following two theorems provide characterizations of 2D peaks
that may be translated into computer programs.
\begin{ther}
We divide 2D peaks into two types, being fractal frames or not,
that can be distinctly characterized as follows. 
\begin{enumerate}
\item $(a,b)$ is a 2D peak and a fractal frame 
\emph{iff}
$(a,b)$ is a fractal frame and $D(a,b)<D_{\text{max}}$. \label{s17}
\item $(a,b)$ is a 2D peak and not a fractal frame
\emph{iff}
$(a,b)$ is a grid frame and
there is a fractal frame $(a',b')$ with $D(a',b')\geq D_{\text{max}}$,
such that $(a',b') \in \Sigma (a,b)$. \label{s18}
\end{enumerate}
\label{s17&18}
\end{ther}
\begin{proof}
(\ref{s17}): Immediate by Defs.~\ref{treetop} and \ref{maxdeep2D}.

(\ref{s18}): If a 2D peak $(a,b)$ is not a fractal frame, then $D(\sigma (a,b))\geq D_{\text{max}}$
by Def.~\ref{maxdeep2D}, so $(a,b)$ is a grid frame by Lemma \ref{s6}.
Applying Lemma \ref{s19}, there is a fractal frame $(a',b') \in \Sigma (a,b)$.
$(a,b)\neq (a',b')$ because one is a fractal frame, the other is not, so
$(a',b') \in \Sigma (\sigma (a,b))$, which by Prop.~\ref{s7}
implies $D(a',b')\geq D_{\text{max}}$.

Conversely, $(a,b)$ is above the diagonal because it is contained in a fractal frame.
$(a,b)\neq (a',b')$ because
$D(a,b)< D_{\text{max}}$ and $D(a',b')\geq D_{\text{max}}$,
implying that $(a,b)$ is not a fractal frame (uniqueness by Lemma \ref{s19})
and not the root frame. 
The other requirements for a 2D peak are established by Lemma \ref{s6}.
\end{proof}
Theorem \ref{s17&18} characterizes all 2D peaks by their relationship to fractal frames.
This is applied in Algorithm 1, that derives all 2D peaks from fractal frames.
However, the next theorem shows that some
2D peaks can be characterized without referring to fractal frames.
\begin{ther}
A nonroot 2D peak has a successor, the depth of which
is either greater or less than $D_{\text{max}}$.
We thus divide 2D peaks into two types,
that can be distinctly characterized as follows. 
Let $(a,b)$ be nonroot. Then
\begin{enumerate}
\item $(a,b)$ is a 2D peak and $D(\sigma (a,b))\geq D_{\text{max}}$
\emph{iff}
$(a,b)$ is a grid frame that is above the diagonal. \label{s13}
\item $(a,b)$ is a 2D peak and $D(\sigma (a,b))<D_{\text{max}}$
\emph{iff}
$(a,b)$ is a fractal frame and there is a grid frame $(a',b')$ that is crossing the diagonal,
such that $(a',b') \in \Sigma (a,b)$. \label{s12}
\end{enumerate}
\label{s12&13}
\end{ther}
\begin{proof}
(\ref{s13}): Immediate by Def.~\ref{maxdeep2D} and Lemma \ref{s6}.
%

(\ref{s12}): If a 2D peak $(a,b)$ has $D(\sigma (a,b))<D_{\text{max}}$,
then $(a,b)$ is a fractal frame by Def.~\ref{maxdeep2D}. Applying Lemma \ref{s9&14}
to $\sigma (a,b)$, there is
a grid frame $(a',b') \in \Sigma (\sigma (a,b))\subset \Sigma (a,b)$.
Frame $(a',b')$ is crossing the diagonal because it contains $\sigma (a,b)$,
which is crossing the diagonal because $(a,b)$ is a fractal frame.

Conversely, $(a,b)\neq (a',b')$ because $(a,b)$ is above the diagonal (a fractal frame)
and $(a',b')$ is crossing the diagonal, and hence $(a',b') \in \Sigma (\sigma (a,b))$.
Since $(a',b')$ is a grid frame, Lemma \ref{s6} gives $D(a',b')<D_{\text{max}}$, which
by Prop.~\ref{s7} implies $D(a,b)\leq D(\sigma (a,b))<D_{\text{max}}$,
and we conclude that $(a,b)$ is a 2D peak. 
\end{proof}
Note: Theorem \ref{s12&13} does not consider the root frame. However,
if the root frame is a 2D peak, then it is of the first type:
a grid frame that is above the diagonal.

It follows from Theorems \ref{s17&18} and \ref{s12&13} that
a 2D peak is either a grid frame, a fractal frame, or both.
The set of 2D peaks $P$ can therefore be divided into three disjoint sets
defined as follows.
$P_{\text{F}}$ are the 2D peaks that are fractal frames only, not grid frames.
$P_{\text{FG}}$ are the 2D peaks that are both fractal frames and grid frames.
$P_{\text{G}}$ are the 2D peaks that are grid frames only, not fractal frames.
Let $G_{\text{a}}$, $G_{\text{b}}$ and $G_{\text{c}}$
be the sets of grid frames that are above, below and crossing the diagonal, respectively.
Let $F_{\text{d}}$ and $F_{\text{s}}$ be the sets of fractal frames that are
deep ($D(a,b)\geq D_{\text{max}}$) and
shallow ($D(a,b)<D_{\text{max}}$), respectively.
In Figs.~\ref{grid}--\ref{2Dpeaks}, all these subsets are illustrated
with different colors. The following corollary summarizes
the relationships between grid frames, fractal frames and 2D peaks:
\begin{coro}
The set of 2D peaks is $P=F_{\text{s}}\cup G_{\text{a}}$.
The intersection between the grid and the fractal is
$P_{\text{FG}}=F_{\text{s}}\cap G_{\text{a}}=F\cap G$.
Furthermore, the 2D peaks can be obtained by the following two expressions,
in which all set unions are between disjoint sets:
\begin{subequations}
\label{FUG&GUF}
\begin{eqnarray}
P &=&
F_{\text{s}}
\bigcup_{(a',b')\in F_{\text{d}}}
G \cap \Delta (a',b'),\label{FUG} \\
&=&
G_{\text{a}}
\bigcup_{(a',b')\in G_{\text{c}}}
F \cap \Delta (a',b').\label{GUF}
\end{eqnarray}
\end{subequations}
\label{korollar}
\end{coro}
\begin{proof}
$P=P_{\text{F}}\cup P_{\text{FG}}\cup P_{\text{G}}$.
Theorem \ref{s17&18} states that $P_{\text{F}}\cup P_{\text{FG}}=F_{\text{s}}$
and that
\[
P_{\text{G}}=
\bigcup_{(a',b')\in F_{\text{d}}}G \cap \Delta (a',b').
\]
Here, $\Delta (a',b')$ is brought into play by Prop.~\ref{s22}.
Theorem \ref{s12&13} states that $P_{\text{FG}}\cup P_{\text{G}}=G_{\text{a}}$
(the root frame would go here) and that
\[
P_{\text{F}}=
\bigcup_{(a',b')\in G_{\text{c}}}F \cap \Delta (a',b'). \qedhere
\]
\end{proof}
Eqs.~(\ref{FUG}) and (\ref{GUF}) outline how the set of 2D peaks
is built up computationally by Algorithm 1 and 2, respectively.
Writing the expressions side by side shows the parallels:
Algorithm 1 takes some fractal frames and then it adds
some grid frames that are contained inside fractal frames.
Algorithm 2 takes some grid frames and then it adds
some fractal frames that are contained inside grid frames.
In both cases, the additional part is the more complicated part, as it
requires searching some forests.
The two Algorithms are algorithmically equivalent in terms of output,
but the transformation in Eq.~(\ref{FUG&GUF}) from $F$-based to $G$-based
facilitates a reduction in execution time, as described in the next section.
\subsection{The fast and exact algorithm}
Algorithm 2 owes its speed to two important ingredients:
One is the grid frame matrix $G$ associated to the parameter $D_{\text{max}}$.
The other is an upper bound associated to the parameter $p_c$.

To compute all bubble stitches of the stitch profile, the algorithm must find
those 2D peaks $(a,b)$ in the bubble context
that have a peak volume
\begin{equation}
p_{\text{v}} (a,b)=\sum_{x\in L(a)}\sum_{y\in L(b)} p_{\text{bubble}} (x,y) 
\end{equation}
that is greater or equal to the probability cutoff $p_c$.
According to Eq.~(\ref{GUF}), one can write an algorithm for obtaining all 2D peaks
using two nested loops that goes through all matrix elements $(a_i,b_j)$
of the grid frame matrix $G$:
If $(a_i,b_j)$ is above the diagonal, it is a 2D peak.
If $(a_i,b_j)$ is crossing the diagonal, a subroutine computes the set
$F \cap \Delta (a_i,b_j)$.
If $(a_i,b_j)$ is below the diagonal, it is skipped.
By piping the resulting frames through a probability cutoff filter, 
we obtain the bubble stitches.

The matrix $G$ is not stored in memory, only the two arrays
$P_{\text{x}}$ and $P_{\text{y}}$ that provide each $a_i$ and $b_j$.
Matrix elements $(a_i,b_j)$ being above, crossing or below the diagonal
refers to the diagonal line in the $xy$-plane, never the diagonal of the matrix.
For each row and column of the matrix there may be zero, one, or more matrix elements
that are crossing the diagonal, as can be seen in Fig.~\ref{grid}.

More specifically, let $G$ be of order $m\times n$
and let the outer loop be over $j=n$ to $1$
and the inner loop over $i=m$ to $1$.
The iteration thus begins at the upper right corner of Fig.~\ref{grid}
and steps along the $y$-axis in the outer loop and the $x$-axis in the inner loop. 
However, we do not have to start at $i=m$ for each $j$.
If $(a_i,b_j)$ is below the diagonal, then $(a_i,b_k)$ is below the diagonal for all $k<j$.
Therefore, we can jump directly to the $i$ that corresponds to the first grid frame
that was not below the diagonal at the previous $j$.
In this way, most of the grid frames that are below the diagonal are
ignored by the algorithm.
While this is a trivial programming trick, we shall now see a less trivial trick,
that ignores most of the grid frames that are above the diagonal.

Recall \cite{bip2003} that the bubble probability is
\begin{equation}
p_{\text{bubble}} (a,b)=\frac{Z_{\text{X10}}(x)\Omega(y-x)Z_{\text{01X}}(y)}{Z}.
\end{equation}
The loop entropy factor $\Omega(y-x)$ is a monotonically decreasing function.
Its largest value in a frame $(a,b)$ is therefore in the lower right corner, i.e.
$\Omega_{\text{max}}=\Omega(y_{\text{start}}(b)-x_{\text{end}}(a))$. Then
\begin{equation*}
p_{\text{v}} (a,b)
\leq 
\frac{\Omega_{\text{max}}}{Z}
\left ( \sum_{x\in L(a)} Z_{\text{X10}}(x) \right )
\left ( \sum_{y\in L(b)} Z_{\text{01X}}(y) \right ),
\end{equation*}
and the bubble peak volume has an upper bound that factorizes.
Using the 1D peak volumes
\begin{subequations}
\begin{eqnarray}
p_{\text{v}}(a) &=& \sum_{x\in L(a)}Z_{\text{X10}}(x)/Z,\\
p_{\text{v}}(b) &=& \sum_{y\in L(b)}Z_{\text{01X}}(y)/Z,
\end{eqnarray}
\end{subequations}
we can write the upper bound as
\begin{equation}
\tilde{p}_{\text{v}} (a,b) =
\Omega(y_{\text{start}}(b)-x_{\text{end}}(a))Zp_{\text{v}}(a)p_{\text{v}}(b).
\end{equation}
If a grid frame $(a_i,b_j)$ has an upper bound below the cutoff,
$\tilde{p}_{\text{v}} (a_i,b_j)<p_c$,
then also $\tilde{p}_{\text{v}} (a_k,b_j)<p_c$
for all $k<i$ for which $p_{\text{v}}(a_k)\leq p_{\text{v}}(a_i)$,
because the loop entropy factor is decreasing.
In that case, their peak volumes are also below the cutoff, of course,
and the algorithm can reject all these frames.

We implement this observation by calculating in advance
the \emph{next bigger goat} $\nbg(i)$ defined by
\begin{enumerate}
\item $p_{\text{v}}(a_k)\leq p_{\text{v}}(a_i)$ for $\nbg(i)<k<i$
\item $p_{\text{v}}(a_{\nbg(i)})> p_{\text{v}}(a_i)$
\end{enumerate}
The $\nbg(i)$ is calculated as follows: A loop
over $i=1$ to $m$ compares each $p_{\text{v}}(a_i)$ successively to
$p_{\text{v}}(a_{i-1})$, $p_{\text{v}}(a_{\nbg(i-1)})$,
$p_{\text{v}}(a_{\nbg(\nbg(i-1))}),\ldots$
until a bigger one is found or the list ends.

For grid frames $(a_i,b_j)$ that are above the diagonal, the algorithm first checks
if $\tilde{p}_{\text{v}} (a_i,b_j)<p_c$, in which case it jumps directly to
$(a_{\nbg(i)},b_j)$.
The $\nbg(i)$ may be undefined,
if there are no bigger $p_{\text{v}}(a_k)$, in which case the inner loop is done
and the outer loop proceeds to the next $j$.
On the other hand, if $\tilde{p}_{\text{v}} (a_i,b_j)\geq p_c$, then the
peak volume has to be calculated and checked.
Although grid frames
may be skipped without having calculated 
neither their peak volumes nor their upper bounds,
the criterion for rejection is exact.
There are no false negatives (or positives).

For each grid frame $(a_i,b_j)$ that is crossing the diagonal, the algorithm
calculates a set of 2D peaks, $F \cap \Delta (a_i,b_j)$,
and checks the peak volume of each.
This set consists of all fractal frames that are contained
inside $(a_i,b_j)$. A mental picture is that $(a_i,b_j)$ must be 
broken into fractal frames (fractured) to avoid crossing the diagonal.
The algorithm searches the subtree $\Delta (a_i,b_j)$
top-down (breadth-first) with a recursive subroutine. A given input frame $(a,b)$
is split into its father frame $\pi (a,b)$ and mother frame $\mu (a,b)$.
Each in turn is then checked as follows:
If it is crossing the diagonal, it is further split by giving it
recursively as input to the subroutine.
If instead it is above the diagonal, it is a fractal frame.
With $(a_i,b_j)$ as input, the subroutine finds $F \cap \Delta (a_i,b_j)$.
(If instead the input is the root frame $(\rho _{\text{x}},\rho _{\text{y}})$,
the subroutine will find all fractal frames $F$.
This was applied in Algorithm 1.)

\begin{figure*}
 \includegraphics[width=13cm]{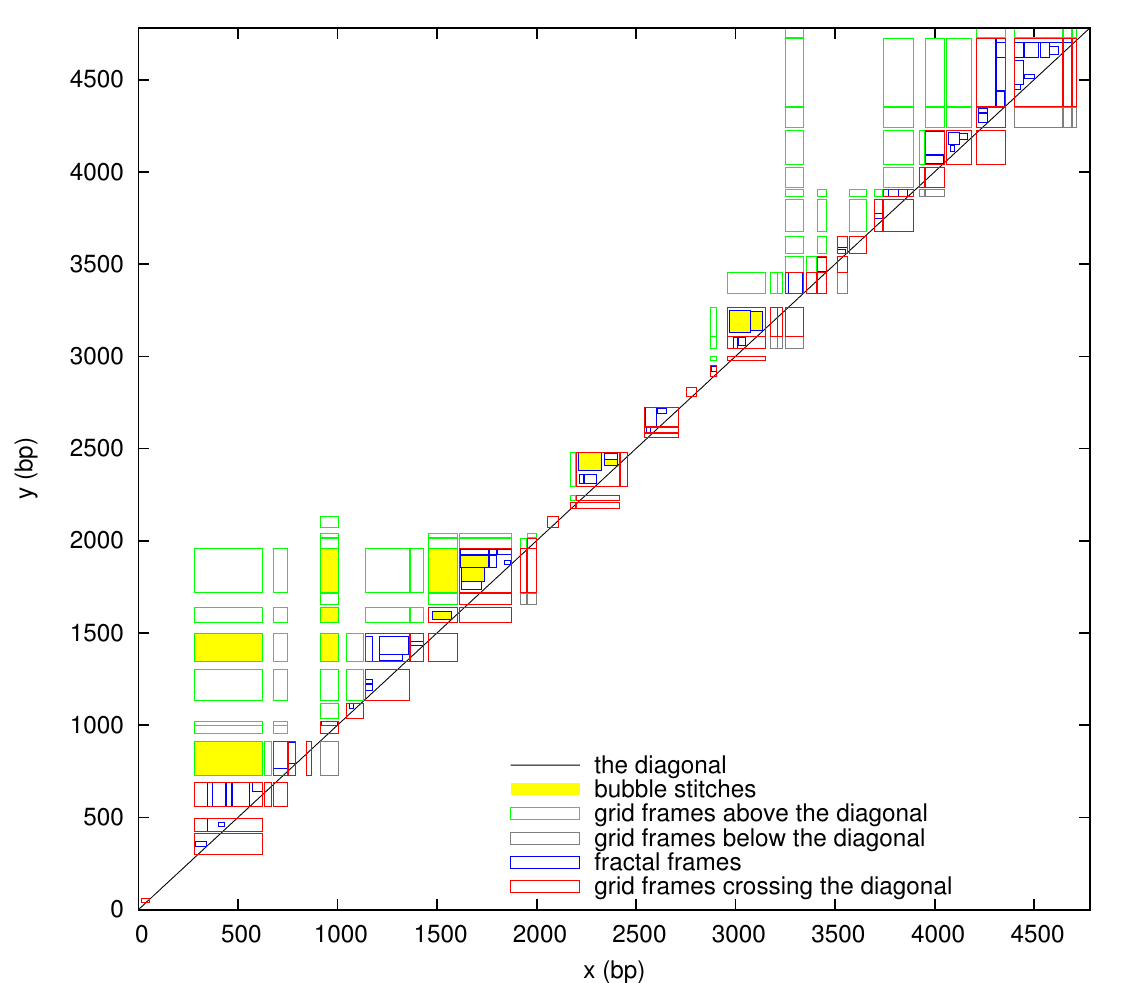}%
 \caption{
The footprints of Algorithm 2 plotted in the $xy$-plane.
These are frames that are visited by Algorithm 2
during its search for the bubble stitches (filled yellow).
The frames are located in a band along the diagonal,
suggesting that the search space is proportional to sequence length.
Grid frames below the diagonal (grey) are skipped.
Grid frames crossing the diagonal (red) are broken into
fractal frames (blue). The bubble stitches (filled yellow) are
those grid frames above the diagonal (green) and fractal frames (blue)
that have $p_{\text{v}} (a,b)\geq p_c$.
Frames with side lengths below 20 bp are not shown to unclutter the figure. 
 \label{footprint}
}
\end{figure*}

Figure \ref{footprint}
shows the resulting search process, by plotting
only frames that are processed by the algorithm,
while the ignored grid frames are blank.
Comparing with Fig.~\ref{grid}, we see that the blank areas
correspond to the great bulk of grid frames both above and below the diagonal,
leaving just an irregular band of frames along the diagonal to be searched.
This is a nice geometric illustration of the reduction
from $O(N^2)$ to $O(N\log N)$ in execution time.
Figure \ref{footprint} also shows that some bubble stitches
are fractal frames contained inside grid frames
that are crossing the diagonal.

The peak volumes $p_{\text{v}}(a,b)$ of some frames must be calculated.
Algorithm 2 spends a considerable fraction of its time on
doing these summations. 
The summation over a bubble frame can be done faster if the frame is big enough,
by exploiting the Fixman-Freire approximation \emph{\`a la} Yeramian
\cite{FF77,Yera1990}.
This does not improve the time complexity, but significantly
reduces the total execution time by some factor.

To compute all helix stitches of the stitch profile, the algorithm
follows exactly the same procedure as described above, but in the helix context.
Eq.~(\ref{GUF}) and the analysis in the previous section applies equally well
to the bubble and the helix contexts.
The various quantities are, of course, replaced by their helix counterparts.
For example, the appropriate diagonal line is applied (Def~\ref{diagonal}).
The main difference is the upper bound on helix peak volume.
Since x and y decouples in the helix probability \cite{tostesen:061922},
\begin{equation}
p_{\text{helix}}(x,y)=\frac{p_{\text{helix}}(x,N)p_{\text{helix}}(1,y)}{p_{\text{helix}}(1,N)},
\end{equation}
we can simply use the peak volume as its own upper bound:
\begin{equation}
\tilde{p}_{\text{v}} (a,b) =
p_{\text{v}} (a,b) =
\frac{p_{\text{v}}(a)p_{\text{v}}(b)}{p_{\text{helix}}(1,N)}.
\end{equation}
The $\Xi (x,y)$ factor \cite{tostesen:061922} is the counterpart of $\Omega (y-x)$, but
an explicit consideration of its monotonicity is not necessary here, because
it is absorbed in the above quantities.
A next bigger goat is then calculated and applied in the same way as for bubbles.
\section{Results and discussion}
\subsection{Time complexity}
By inspection of Algorithm 2, we observe that it visits
at least $O(N)$ and at most $O(N^2)$ matrix elements of $G$.
Furthermore, it performs sorting, which is known to scale as $O(N\log N)$.
The time complexity is therefore
between $O(N\log N)$ and $O(N^2)$.
The execution time depends on the fraction of ignored grid frames
above the diagonal, which depends on
the specific sequence, temperature, and other input parameters.
A theoretical analysis of these dependencies is complicated.

\begin{figure}
 \includegraphics[width=8.5cm]{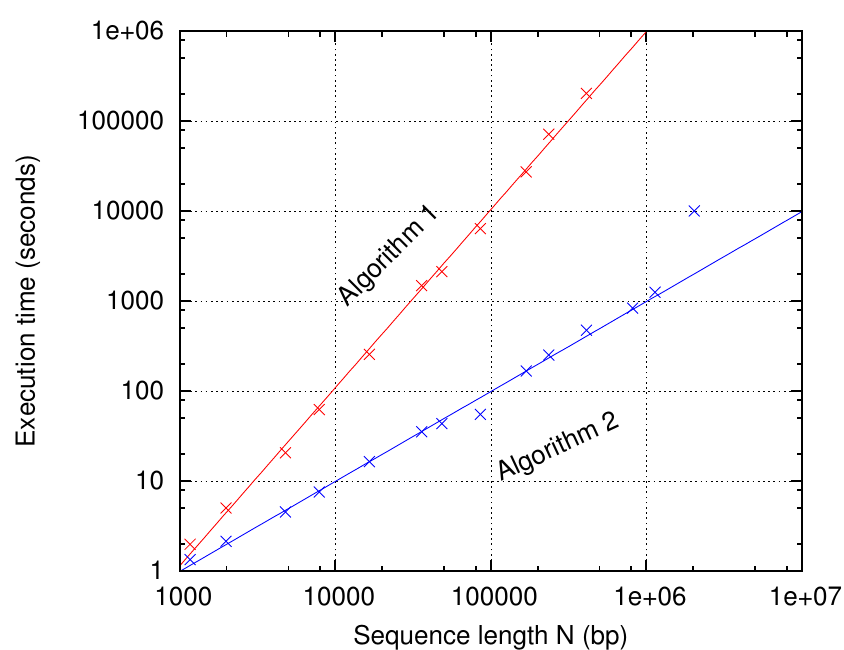}%
 \caption{
Algorithm 1 is quadratic and Algorithm 2 is linear.
The log-log plot shows the execution time versus sequence length
of Algorithm 1 (red) and Algorithm 2 (blue).
The straight lines are fits to the data points
with slopes $1.97955 \pm 0.02923$ (red)
and $0.99953 \pm 0.02016$ (blue). 
 \label{speed1}
}
\end{figure}

Empirical testing of the execution times were done instead,
using a test set of 14 biological sequences
with lengths selected to be evenly spread on a log scale spanning three decades.
A minimum length of 1000 bp was required.
Most of the test sequences are genomic sequences,
so as to represent the typical usage of the algorithm.
The sequence lengths and accession numbers are:
\begin{itemize}
\item 1168 bp [GenBank:BC108918]
\item 1986 bp [GenBank:BC126294]
\item 4781 bp [GenBank:BC039060]
\item 7904 bp [GenBank:NC\_001526]
\item 16571 bp [GenBank:NC\_001807]
\item 36001 bp [GenBank:AC\_000017]
\item 48502 bp [GenBank:NC\_001416]
\item 85779 bp [GenBank:NC\_001224]
\item 168903 bp [GenBank:NC\_000866]
\item 235645 bp [GenBank:NC\_006273]
\item 412348 bp [GenBank:AE001825]
\item 816394 bp [GenBank:NC\_000912]
\item 1138011 bp [GenBank:AE000520]
\item 2030921 bp [GenBank:NC\_004350]
\end{itemize}
The algorithms were written in Perl and run on a
Pentium 4, 2.4 GHz, 512 KB cache, 1 GB memory,
PC with Linux (CentOS).
In Fig.~\ref{speed1},
the speeds of Algorithms 1 and 2 are compared.
Algorithm 2 is orders of magnitude faster
than Algorithm 1 for sequences longer than 100 kbp.
While all the 14 sequences were computed by Algorithm 2,
the three longest sequences were aborted
by Algorithm 1, because of too long execution times.
To ensure that the computational tasks were comparable,
all sequences were computed at their melting temperatures $T_{\text{m}}$,
rather than one temperature for all,
such that all sequences had the same fractions of helical regions and bubbles.
For both algorithms, straight lines were fitted to the data in the log-log plot.
For Algorithm 2, however, the longest sequence (2 Mbp) is considered
an outlier and thus excluded from the fit. 
This sequence's execution time was overly increased, because the required memory
exceeded the available 1 gigabyte RAM.
For Algorithm 1, the slope of the fit is $1.97955 \pm 0.02923$,
suggesting that it has time complexity $O(N^2)$.
For Algorithm 2, the slope of the fit is $0.99953 \pm 0.02016$.
This is interpreted as the time complexity $O(N\log N)$, but
with the logarithmic component being too weak to distinguish
$O(N\log N)$ from $O(N)$.

\begin{figure}
 \includegraphics[width=8.6cm]{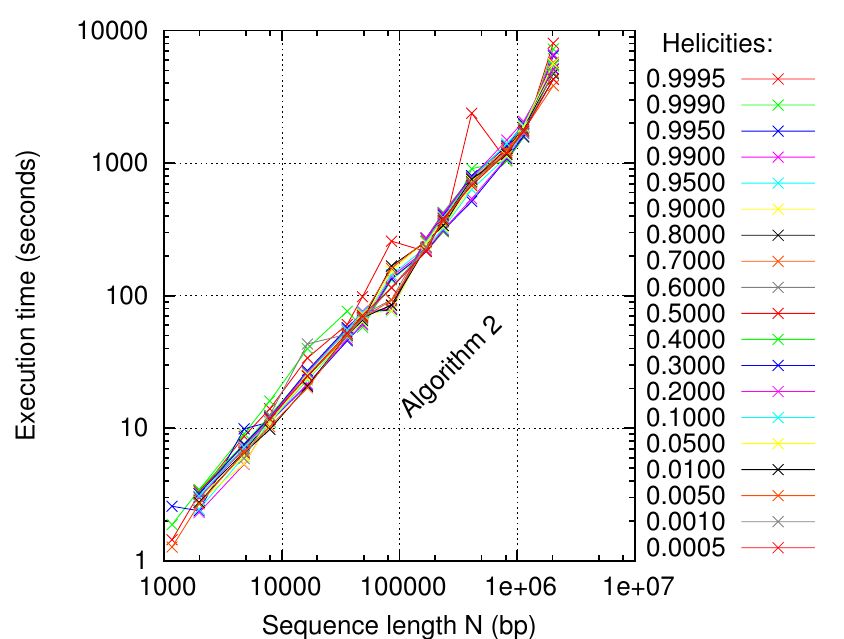}%
 \caption{
Algorithm 2 is fast at all temperatures.
The total execution times are plotted versus
sequence length for each of the listed helicity values. 
 \label{speed2}
}
\end{figure}

The execution time of Algorithm 2 is just as much a property
of the underlying energy landscape depending on the input,
as it is a property of the algorithm.
Could it be that other input parameters and/or sequences
than was used in Fig.~\ref{speed1}---say,
away from the melting points---would exhibit the time complexity $O(N^2)$?
Figure \ref{speed2}
shows the speed of Algorithm 2 over the whole melting range
of temperatures.
Each sequence in the test set was computed
at temperatures corresponding to the helicity values:
0.9995, 0.999, 0.995, 0.99, 0.95, 0.9, 0.8, 0.7,\dots, 0.2,
0.1, 0.05, 0.01, 0.005, 0.001, and 0.0005.
This helicity range approximately corresponds to the temperature range
$T_{\text{m}} \pm 10^{\circ }\text{C}$
and it covers most of the melting transitions.
Although the curves for the individual helicity values may not
be easily distinguished in Fig.~\ref{speed2}, it appears that
all curves have similar slopes and that they are close to each other,
i.e., the variation in execution time is below $50\%$.
This indicates that the helicity (or temperature) value
has only a small influence on the total execution time.
The time complexity $O(N\log N)$ seems to be robust.

\begin{figure}
 \includegraphics[width=8.6cm]{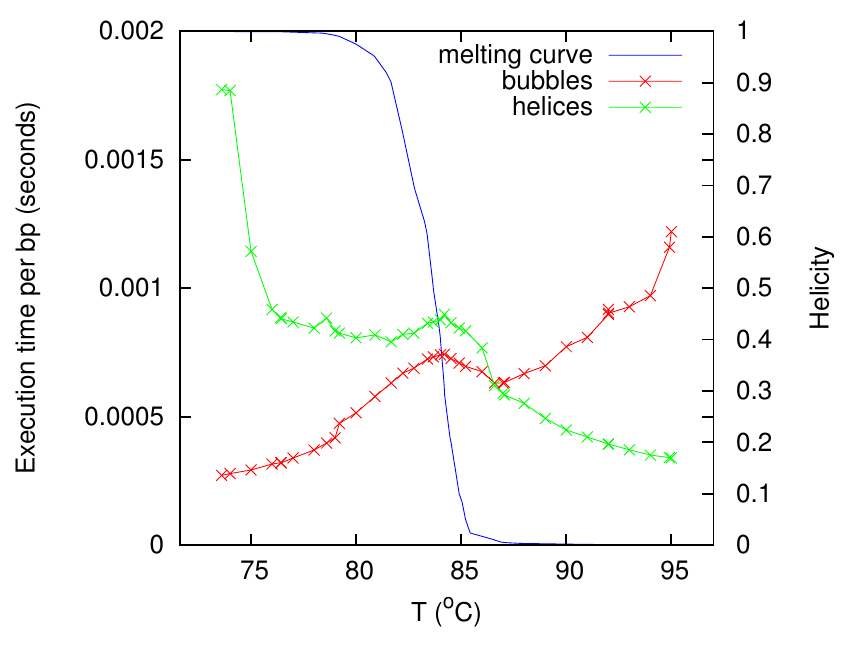}%
 \caption{
Bubble and helix execution times versus temperature.
For the sequence [GenBank:NC\_001807], the bubble time (red) and helix time (green)
divided by sequence length (16571 bp) is plotted versus $T$.
The melting curve (blue) shows the helicity $\Theta$
(on the right vertical axis) as a function of $T$,
indicating the melting midpoint: $\Theta=0.5$ at $T_{\text{m}}=83.7^{\circ }\text{C}$.
\label{speed3}
}
\end{figure}

\begin{figure}
 \includegraphics[width=8.6cm]{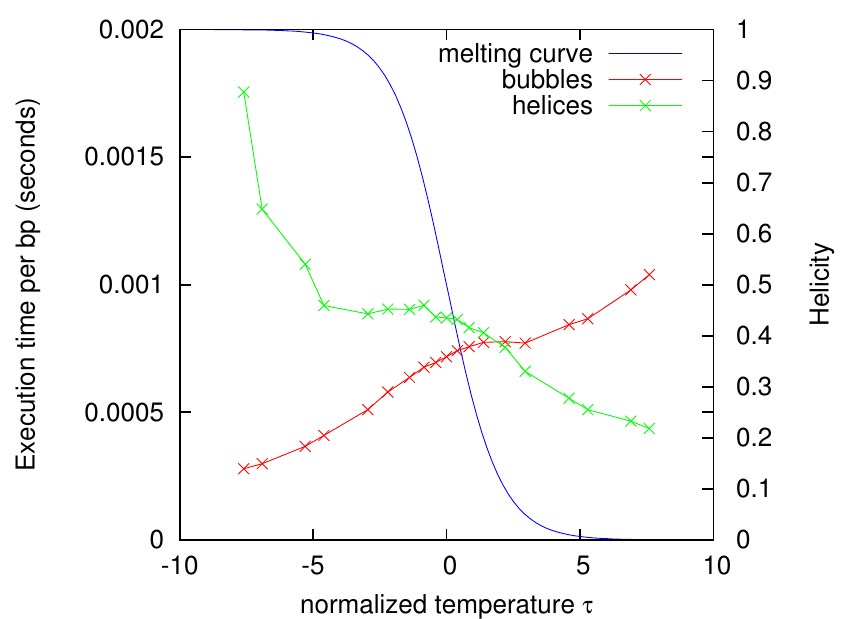}%
 \caption{
Sequence-averaged bubble and helix execution times.
The bubble time per basepair (red) and helix time per basepair (green)
averaged over all sequences are
plotted versus the normalized temperature $\tau$
for each of the helicity values listed in Fig.~\ref{speed2}.
The melting curve (blue) shows the helicity $\Theta$
(on the right vertical axis) as a function of $\tau$,
indicating the melting midpoint: $\Theta=0.5$ at $\tau=0$.
\label{speed4}
}
\end{figure}

However, a stronger temperature dependence is revealed
when considering the computations of bubble stitches and helix stitches
separately. Two independent subroutines of Algorithm 2
compute the bubble stitches and the helix stitches,
both following the procedure outlined in the previous section.
The rest of Algorithm 2's computation,
including the initial computation of at least four partition function arrays
\cite{bip2003},
is called the overhead.
Correspondingly, the total execution time $t_{\text{total}}$ is the sum of the
bubble execution time $t_{\text{bubble}}$, the helix execution time $t_{\text{helix}}$,
and the overhead execution time $t_{\text{overhead}}$.
By simply switching off the bubble subroutine (i.e. $t_{\text{bubble}}=0$)
and measuring the total execution time, we obtain $t_{\text{helix}}+t_{\text{overhead}}$.
Likewise, by switching off the helix subroutine, we measure $t_{\text{bubble}}+t_{\text{overhead}}$.
In the following, we refer to $t_{\text{bubble}}+t_{\text{overhead}}$ as the bubble time
and $t_{\text{helix}}+t_{\text{overhead}}$ as the helix time.
As an example, Fig.~\ref{speed3} shows the results for the 16571 bp [GenBank:NC\_001807].
The bubble and helix times are divided by sequence length and plotted
as a function of temperature.
Both of them have clearly a strong temperature dependence.
The melting curve is also plotted in Fig.~\ref{speed3},
indicating that most of the melting occurs in the temperature range
$80$--$85^{\circ }\text{C}$.
Plots like Fig.~\ref{speed3} were made for each sequence
in the test set, but the average behavior is more interesting.
To average times of the order $O(N)$ over sequences of different lengths,
one should divide them by sequence length as in Fig.~\ref{speed3}.
However, to plot as a function of temperature would not be meaningful,
because the sequences have different $T_{\text{m}}$'s and different melting ranges.
On the horizontal axis, instead, we use a normalized temperature,
\begin{equation}
\tau=\log(\frac{1-\Theta}{\Theta}),
\end{equation}
defined such that the melting curve becomes a sigmoid:
\begin{equation}
\Theta=\frac{1}{1+\exp(\tau)}.
\end{equation}
For each $\tau$-value (or equivalently for each $\Theta$-value),
the bubble times and helix times divided by sequence length 
averaged over all sequences are plotted in Fig.~\ref{speed4}.
The curves have a similar temperature dependence as in Fig.~\ref{speed3}.
The helix time decreases monotonically
(except for a shoulder), while 
the bubble time increases monotonically
(except for a shoulder). Both of them have an about four-fold
difference between their maximum and minimum. 
Qualitatively, the curves are kind of mirror symmetric, but the helix time
is generally greater than the bubble time,
the two curves cross each other at $\Theta=0.12$.
It seems that adding the two curves would give a more or less horizontal curve,
i.e., the total execution time has much less temperature dependence. 

We may understand this interchange between bubble time and helix time
in terms of the melting process.
If we assume that the bubble time is proportional
to the area of the footprint in Fig.~\ref{footprint}, and that this
is proportional to the average length
of potential bubbles at that temperature, then we would
expect the bubble time to increase with temperature,
because bubbles grow as DNA melts.
Likewise, we would expect the helix time to decrease with temperature, 
because helical regions diminish as DNA melts.

In this article, Blake \& Delcourt's parameter set \cite{BD98}
as modified by Blossey \& Carlon \cite{PhysRevE.68.061911}
was used with $[\text{Na}^{+}]=0.075\text{ M}$.
The maximum depth and probability cutoff parameters were
$D_{\text{max}}=5$ and $p_{\text{c}}=0.01$ in Figs.~\ref{grid}--\ref{footprint},
$D_{\text{max}}=3$ and $p_{\text{c}}=0.02$ in Fig.~\ref{speed1}, and
$D_{\text{max}}=3$ and $p_{\text{c}}=0.0001$ in Figs.~\ref{speed2}--\ref{speed4}.
The sequence [GenBank:BC039060] was used for producing
Figs.~\ref{p1}--\ref{footprint}.
A systematic test of how the execution time depends on
$D_{\text{max}}$ and $p_{\text{c}}$ has not been performed.
\subsection{Discussion}
For an algorithm to be called efficient, it should solve the task at hand
with optimal time complexity. It should not introduce approximations,
that would just amount to a reformulation of a simpler, but different task.
In this study, the task is to compute a stitch profile
based on the Poland-Scheraga model with Fixman-Freire loop entropies.
With this model, the time complexity must be at least $O(N\log N)$.
Indeed, this is achieved by Algorithm 2 under a wide range of conditions.
Algorithm 2 does not acquire a speedup by any \emph{windowing} approximation,
by which the sequence would first be split into smaller independent sequences.
Neither does it rely on limiting the problem to a maximal bubble length.
Therefore, Algorithm 2 is efficient.
In computational RNA and protein studies, a maximal loop size
is sometimes imposed as a heuristic
for reducing time complexity by one order.
Similarly, a maximal DNA bubble size of 50 bp has been reported
in computations of low temperature bubble probabilities
in the Peyrard-Bishop-Dauxois model
\cite{erp:218104}.
In contrast, Algorithm 2 can find bubbles of whatever size at any temperature.
In the 48502 bp [GenBank:NC\_001416], for example,
bubbles and helical regions may be up to around 20000 bp long
\cite{lambdastitch}.
Although Algorithm 2 has no explicit notion of a maximal bubble length,
it may implicitly detect length limitations for both bubbles
and helical regions by the absence of the ``next bigger goat''.
In this way, Algorithm 2 can adapt to the input sequence.
This adaptation is evident in Fig.~\ref{speed4},
where the bubble execution time grows
as bubbles get bigger at higher temperatures. Conversely, the helix execution
time decreases as the helical regions gradually melt away.

However, the time complexity was not proven to be $O(N\log N)$ under all conditions.
It is still an open question whether there is a transition to
time complexity $O(N^2)$ in some peripheral regions of the input parameter space.
But based on results so far, a fast computation would be expected in most situations.

How fast is Algorithm 2?
Figures \ref{speed1} and \ref{speed2} show that
the Perl implementation runs on an old desktop PC
at the speed of roughly 1000 basepairs per second.
With today's computers, assuming twice that speed and enough memory,
the \emph{E. coli} genome would take 39 minutes,
the yeast genome would take 1.7 hours,
and the largest human chromosome would take 35 hours.
In some types of low temperature melting studies,
the features of interest are the bubbles
rather than the helical regions.
In such applications, switching off the computation of helix stitches
can speed up the algorithm several times.
As Fig.~\ref{speed4} indicates, the helix time is about twice the
bubble time at helicity equal to 0.95, that is,
the speedup would be about threefold.
The largest human chromosome would be done in ten hours.
On a computer cluster, the human genome could be computed
in a day.
Such bubbles could then be compared to TFBS, TSS, replication origins,
viral integration sites, etc.

However, the required memory grows with sequence length and
for sequences longer than 2 Mbp, more than 1 GB was needed.
The memory usage has not been tested further and
the space complexity has not been discussed in this article.
Some memory optimization of the Perl implementation must be done
before such test can reflect the space complexity.
While the algorithm is efficient in terms of time complexity, 
the code has room for optimization of both speed and memory usage.
However, the space complexity is believed to be $O(N)$, which means that
the algorithm would eventually become out of memory for long enough sequences.
A standard solution is to introduce efficient use of disk space instead,
which could reduce the memory usage to $O(1)$, without
increasing the time complexity.
\section{Conclusions}
The fast algorithm described in this article
enables the computation of stitch profiles
of genomic sequences.
Melting features of interest,
such as bubbles, helical regions, and their boundaries,
are computed directly, rather than relying
on visualization or educated guesses.
The algorithm is exact. It does not achieve its speed by
approximations, such as windowing or maximal bubble sizes.
Genomewide comparisons of bubbles with TSS, replication origins,
viral integration sites, etc., are proposed.
The algorithm is available in Perl code from the author.
Online computation of stitch profiles is available on our web server,
which has recently been upgraded to run Algorithm 2
\cite{narweb,stitchprofiles.uio.no}.
\section{Competing interests}
The author(s) declare that they have no competing interests.
\begin{acknowledgments}
Discussions with Eivind Hovig, Geir Ivar Jerstad, and Torbj\o rn Rognes
on genome browsers gave the impetus to this work.
Funding to this work was provided by
FUGE--The national programme for research in functional genomics in Norway.
\end{acknowledgments}

\bibliography{stitchintime}

\end{document}